# Analytical models of icosahedral shells for 3D optical imaging of viruses

Aliakbar Jafarpour

*Department of Biomolecular Mechanisms, Max-Planck Institute for Medical Research, Jahnstraße 29, 69120 Hiedelberg, Germany*
*jafarpour.a.j@ieee.org*

**Abstract**: A *modulated icosahedral shell with an inclusion* is a concise description of many viruses, including recently-discovered large double-stranded DNA ones. Many X-ray scattering patterns of such viruses show major polygonal fringes, which can be reproduced in image reconstruction with a homogeneous icosahedral shell. A key question regarding a low-resolution reconstruction is how to introduce further changes to the 3D profile in an efficient way with only a few parameters. Here, we derive and compile different analytical models of such an object with consideration of practical optical setups and typical structures of such viruses. The benefits of such models include 1) inherent filtering and suppressing different numerical errors of a discrete grid, 2) providing a concise and meaningful set of descriptors for feature extraction in high-throughput classification/sorting and higher-resolution cumulative reconstructions, 3) disentangling (physical) resolution from (numerical) discretization step and having a vector graphics format for visualization or further analysis at arbitrary scales, 4) eliminating the phase-retrieval step and enforcing transparent, relevant, and controlled type/level of *a-priori* information in a real-space formulation, and 5) evaluating the reflections and surface resonances of an icosahedral object, and hence corrections for the common scattering model.

## *1. Introduction*

### 1.1. Biophysical context

Large double-stranded DNA viruses (1) are interesting entities not only at fundamental biological level (a virus being infected by another virus, complex correlated genetic and proteomic codes …), but also as test cases for single-particle 3D imaging with X-rays or electron microscopy (EM).

Major dark-bright polygonal fringes observed in experimental 2D X-ray scattering patterns (2) can be reproduced with just a homogeneous icosahedral shell. A more detailed reconstruction also includes A) modulations of the shell and B) an inclusion inside the shell. These two considerable modifications (in real-space) introduce small perturbations to the dominant polygonal fringes (in the Fourier space). Disentangling such weak/noisy/correlated signatures can be facilitated by an analytical model.

Here, we are concerned with translating the known geometrical information (icosahedral shell) and unknown information (shell gradients and inclusion) into an efficient few-parameter analytical model of scattering. We demonstrate the practical feasibility of such minimal models in a follow-up report by applying them to experimental X-ray scattering data from two viruses, namely CroV (3) and PBCV-1 (4).

Extraction of the highest possible resolution or a comprehensive validation of the reconstructions is beyond the scope of this contribution. The insight from this study, however, can be used to define tangible figures of merit for validation purposes.

While the mathematical derivations regarding Fourier transforms are self-consistent, applying them to optical scattering data is based on the common (Geometrical) scattering model. General electromagnetic considerations regarding this assumption have been detailed elsewhere (5,6). Here, we address two such challenging cases (reflection and resonance) for the specific case of an icosahedral object.

### 1.2. Mathematical context and the proposed approach

Despite a rich literature on icosahedral symmetry, considerably less attention has been paid to applying such formulations to the icosahedron itself. Only recently, an analytical formula for the 3D Fourier transform of a solid icosahedron has been reported (7). Previous formulations of icosahedral symmetry are dominantly concerned with either

1) A quasi-crystal as the projection of a 6D object with translational symmetry in 3D space (8), or
2) Low-resolution fit to the solution scattering data with a model comprising icosahedrally-symmetric clusters of dense atoms *on a spherical shell* (9), or
3) General expressions (with a lot of parameters and potential uncertainties about uniqueness of solutions) for general icosahedrally-symmetric objects (10,11).

Our approach here has nothing to do with the first one (quasi-crystals), even though such a viewpoint may give additional insights. The second approach (spherical surfaces) is indeed a 2-variable problem, whereas the approach proposed here models the full 3D *volume* density by considering flat surfaces of a solid icosahedron (with possible deformations). The approach proposed here can easily evolve into the third one and achieve its acclaimed completeness by choosing a thick enough icosahedron and sufficient number of modulating terms. However, our major concern (in handling low-resolution X-ray scattering data) and the major benefit of the approach proposed here is explicit and controlled use of *a-priori* information and using as few nontrivial parameters as possible, as depicted in Figure 1.

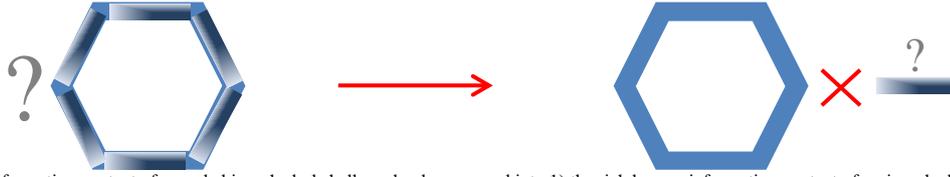

**Figure 1**: The rich information content of a graded icosahedral shell can be decomposed into 1) the rich known information content of an icosahedral shell with uniform density and 2) one few-parameter *modulating* function.

**1.3. Outline**

This report has been organized as follows. The geometry of an icosahedron and the rotations associated with it are briefly defined in Section 2. Analytical and semi-analytical formulae for the 3D Fourier transform of an icosahedral object with different modulation schemes (mimicking apparent patterns in viruses and having simple mathematical representations) are presented in Section 3. Section 4 addresses the topic of spherical harmonic spectrum of an icosahedral object and its relation with Fourier spectrum. Section 5 presents analytical formulation of reflections from icosahedral objects; required for systematic optical alignment and interpretation of unorthodox scattering patterns from stream of hydrated viruses. Finally, conclusions are made in Section 6, and mathematical derivations are detailed and further discussed in Sections I to XIV of *Supplementary Materials*.

## *2. Geometrical model of an icosahedron*

### 2.1. Coordinates and orientation

A solid icosahedron with uniform *density* ($\rho=1$) embedded in a zero-density environment is characterized with a single geometrical parameter of *size* $R_{ico}$, taken to be the radius of the corresponding circumscribed sphere (one that touches all vertices of the icosahedron). Without loss of generality, we consider the case with $R_{ico}=1$ here, and scale the results for arbitrary cases later. The coordinate system is defined such that

1. The z-axis is along a 5-fold symmetry axis of the icosahedron.
2. The origin of the coordinate system coincides with the center of the icosahedron.
3. One of the five "upper" vertices (with $0 < z < 1$) lies in the x-z plane.

With this common convention and defining the angles $\Omega = 2\pi/5$, $\theta_c = \tan^{-1}(2)$, and $\alpha = \tan^{-1}(3 - \sqrt{5})$, the coordinates and the indices of the 12 vertices (Figure 2 (left)) in *spherical* coordinate $(r, \theta, \varphi)$ can be written as:

– Top vertex: $V_1$ (1,0,0)
– 5 Upper vertices: $V_{n+1}$ $(1, \theta_c, (n-1)\Omega)$, where $1 \leq n \leq 5$
– 5 Lower vertices: $V_{n+6}$ $(1, \pi - \theta_c, (n+1.5)\Omega)$, where $1 \leq n \leq 5$
– Bottom vertex: $V_{12}$ $(1, \pi, 0)$.

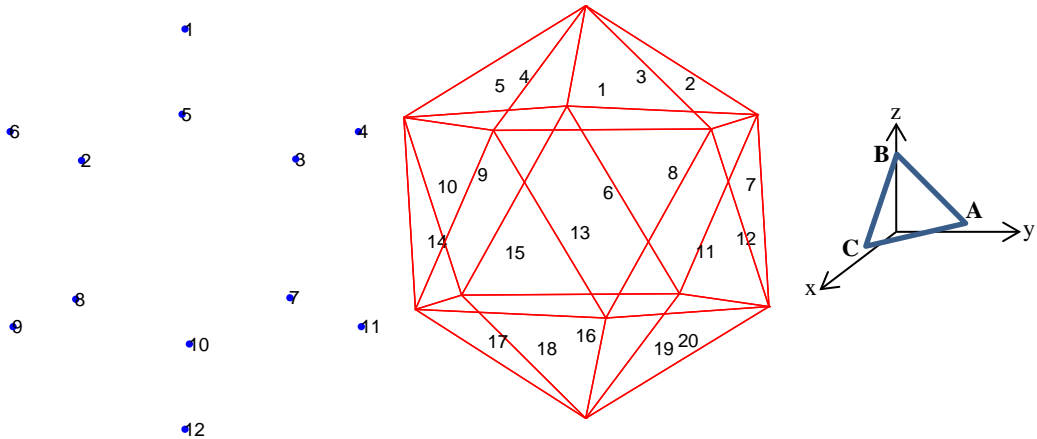

**Figure 2**: (Left) Indexed vertices and (Middle) indexed faces of an icosahedron with the face number written *in the middle* of a face. (Right) The face number 1 of the icosahedron is formed by the vertices ($V_1, V_2, V_3$), which are also denoted as (B,C,A). The three vertices B, C, and A define not only the *flat* triangle BCA, but also a *spherical* triangle BCA, which is 1/20[th] of the unit sphere. With icosahedral symmetry, the spherical map of 3D rotations (excluding mere in-plane ones) is limited to 1/60[th] of the surface of the sphere. This irreducible rotation zone is 1/3[rd] of the surface of the *spherical* triangle BCA.

The lower half of the icosahedron is the *inversion* of the upper half. With vertex indices defined as above, one can write in the *Cartesian* coordinate: $V_{n+6} = -V_n$ for $1 \leq n \leq 5$ and $V_{12} = -V_1$.

By connecting *adjacent* vertices $\{V_i, V_j, V_k\}$, the convex surface of the icosahedron is formed as a set of 20 equilateral triangles, shown in Figure 2 (middle). The coordinate of the mid-point of the triangle $V_i V_j V_k$ is the arithmetic mean of vertex coordinates in the *Cartesian* system $(OV_i + OV_j + OV_k)/3$, and has a distance of $\cos(\alpha)$ from the origin. Specifically, the midpoint of the triangle $V_1V_2V_3$, shown as BCA in Figure 2 (right), has the following *spherical* coordinates $F = (\cos(\alpha), \alpha, \Omega/2)$.

## 2.2. Rotations

As a *3D object*, an icosahedron has 2-fold, 3-fold, and 5-fold symmetry for planar rotations (around specific directions) and overall 60-fold symmetry for 3D rotations. A planar *2D section* through an icosahedron can feature 2-fold, 3-fold, or 5-fold symmetry. An orthogonal *2D projection* of an icosahedron can feature a combined (such as 6-fold) symmetry, as frequently seen in X-ray scattering or EM measurements.

There exist 60 rotation matrices that superimpose an *ordered* triangle (triplet of adjacent faces) $V_iV_jV_k$ on another ordered triangle $V_mV_nV_p$ (including itself). These rotation matrices are $R_{60}([i,j,k] \rightarrow [m,n,p]) = [v_m | v_n | v_p] [v_i | v_j | v_k]^{-1}$, where $v_x$ is the (3x1) *column* matrix representation of the coordinate vector $V_x$.

Only 1/3 of these rotations (20 ones) superimpose one face onto all faces (including itself). The other 2/3rds simply rotate the triangles around their centers and change the sequence of vertices. We consider $R_{20}$ to be all 20 rotations that superimpose $V_1V_2V_3$ onto all faces. These 20 rotation matrices are not unique. However, given the 3-fold symmetry of a triangle, applying such rotation matrices (irrespective of the *permutation* of vertex indices) reproduces all 20 copies of the triangle $V_1V_2V_3$. In derivation of Fourier transforms, both $R_{60}$ and $R_{20}$ matrices are required (Section I in *Supplementary Materials*).

## 2.3. Irreducible rotation zone

In the Euler angles model of 3D rotation, two (polar and azimuthal) angles can be represented on the surface of a sphere. The implication of (mere) icosahedral symmetry for 3D rotations is to limit the surface of this sphere to 1/60[th], referred to as the irreducible rotation zone (**I**). One possible choice of this irreducible zone (as in the common EM data processing suite EMAN) is the spherical triangle $BCF'$, where $F' = (1, \alpha, \Omega/2)$ is the midpoint of the spherical triangle BCA. So, zone **I** is characterized as : $0 \leq \phi \leq \Omega/2$ ; $0 \leq \theta_{CA} \leq \cot^{-1}(G\cos(\Omega - \phi))$, in which $G = (\sqrt{5}+1)/2$ is the golden ratio (12).

With additional inversion (Friedel) symmetry, the irreducible rotation zone is further limited by one half (**I$_h$**); i.e., to 1/120[th] of the surface of the sphere. The set of all possible 2D (EM or X-ray) snapshots of a 3D object with icosahedral symmetry (excluding redundant in-plane rotations) is sufficiently described with 2 rotation parameters limited to the irreducible rotation zone (**I** or **I$_h$**). An **I$_h$** map of 2D projections of the bovine papilloma virus (with different, yet equivalent coordinates of the irreducible zone) is shown in Figure 12 in (13).

## *3. 3D Fourier spectrum*

### 3.1. Solid icosahedron with uniform density

Each triangular face of an icosahedron caps a volume seen at a solid angle of $4\pi/20$ stradians from the origin. This volume is further capped with 3 other triangular faces; i.e., a total of 4 faces making an *irregular* tetrahedron. Knowing the 3D Fourier transform of one such tetrahedron (and utilizing the commutative property of 3D rotation and 3D Fourier operators), the 3D Fourier transform of the icosahedron is obtained by applying the 20 rotations $R_{20}$ and adding up the results. So, the problem (not only for an icosahedron, but also for other symmetric convex polyhedra) is simply reduced to the Fourier transform of a single tetrahedron.

The first tetrahedron, OBCA in Figure 2 (Top-Right), is limited to the following coordinates (Section II in *Supplementary Materials*):

*Spherical coordinate:*
$$0 \leq \phi \leq \Omega$$
$$0 \leq \theta_{CA}(\phi) \leq \cot^{-1}[\cot(\theta_c)\sec(\Omega/2)\cos(\phi - \Omega/2)]$$
$$0 \leq r_{BCA}(\theta, \phi) \leq [\cos(\theta) + \sin(\theta)\tan(\alpha)\cos(\phi - \Omega/2)]^{-1}$$

*Cartesian coordinate:*
$$0 \leq y \leq \sqrt{G}$$
$$y/\tan(\Omega) \leq x(y) \leq \sin(\theta_c) - \tan(\Omega/2)y$$
$$\cot(\theta_c)[x + \tan(\Omega/2)y] \leq z(x, y) \leq 1 - \tan(\alpha)\cos(\Omega/2)[x + \tan(\Omega/2)y]$$

### 3.1.1. Integration in the Cartesian coordinate

Knowing the boundaries of the first tetrahedron, its Fourier transform is written as

$$F_{T1/20-S-IH}(\mathbf{q}) = \int_{y=0}^{y=c_2} dy e^{-i2\pi y q_y} \int_{x=y/tan(\Omega)}^{x=sin(\theta_c)-tan(\Omega/2)y} dx e^{-i2\pi x q_x} \int_{z=cot(\theta_c)x+c_6 y}^{z=1+c_1 x + c_1 tan(\Omega/2)y} dz e^{-i2\pi z q_z}$$

The constants $c_n$ are related to trigonometric functions of the angle $\Omega = 2\pi/5$, and have been defined in Section III of *Supplementary Materials*. They have all been listed with reduced irrational expressions that can be calculated with arbitrary floating point accuracy.

All integrals encountered in the calculation of $F_{T1/20-S-IH}(\mathbf{q})$ simply include exponential integrands (possibly modulated with a monomial). Using a simple lemma, defining a complex generalization of the *sinc* function, and some simple functions of the spatial frequency vector **q**, such integrals can be easily evaluated analytically (Sections IV, V, and VI of *Supplementary Materials*). Evaluation of the first, the second, and the third integrals in different cases have been detailed in Sections VII, VIII, and IX of *Supplementary Materials*, respectively. The final result is

$$F_{T1/20-S-IH}(\boldsymbol{q}) = [\eta_0 + \eta_1]E(q_y - c_{13}q_x) + \eta_2 E(q_y + \eta_4) + \eta_3 E(q_y + \eta_5)$$

, where

$$E(p) = \int_{y=0}^{y=c_2} e^{-i2\pi py} dy = \begin{cases} c_2 \dfrac{e^{-i2\pi c_2 p} - 1}{-i2\pi c_2 p} & if\ p \neq 0 \\ c_2 & if\ p = 0 \end{cases}$$

$$\eta_0(q_x, q_z) = \frac{-e^{-i2\pi(q_z + \sin(\theta_c)(q_x + c_1 q_z))}}{4\pi^2 q_z (q_x + c_1 q_z)}, \eta_2(q_x, q_z) = \frac{e^{-i2\pi q_z}}{4\pi^2 q_z (q_x + c_1 q_z)}, \eta_4(q_x, q_z) = c_{14}q_x + +c_1 c_4 q_z$$

$$\eta_1(q_x, q_z) = \frac{e^{-i\pi \sin(\theta_c)(2q_x + q_z)}}{2\pi^2 q_z (2q_x + q_z)}, \eta_3(q_x, q_z) = \frac{-1}{2\pi^2 q_z (2q_x + q_z)}, \eta_5(q_x, q_z) = \cot(\Omega)\,q_x + (c_4/2)q_z$$

### 3.1.2. Ambiguities

While the ambiguity of the function $E(p)$ at $p = 0$ has been explicitly removed, vanishing denominators of other terms give rise to $\infty - \infty$ ambiguities. To be on the safe/fast side computationally and also to have a closed-form expression (rather than an infinite series) analytically, we have treated the calculation of the triple integral in such cases separately (Section IX of *Supplementary Materials*). As expected, all ambiguities correspond to a finite value and not a singularity.

### 3.1.3. Full icosahedron

Given the inversion symmetry of an icosahedron, the Fourier transform of the *entire* object is real, and has the same contribution (to the real part) from the upper and the lower tetrahedrons:

$$F_{S-IH}(\boldsymbol{q}) = \sum_{n=1}^{20} F_{T1/20-S-IH}(R_n^{-1}\boldsymbol{q}) = 2\sum_{n=1}^{10} Re\{F_{T1/20-S-IH}(R_n^{-1}\boldsymbol{q})\}$$

### 3.1.4. Integration in Spherical coordinate

Having evaluated the Fourier transform triple integral in the Cartesian coordinate, one can establish different identities by rewriting the Fourier transform in spherical coordinate. Some such identities have been listed in Sections X and XI of *Supplementary Materials*.

## 3.2. Achiral spherical deformation

### 3.2.1. Significance

An important modulation of a solid icosahedron is one that deforms its surface towards a spherical shell with icosahedral modulation. Such a deformation can be quantified using a single parameter of *sphericity*, (*sphere factor* in the software suite Chimera (14)). Mathematically, this deformation could also be modeled with the modulation schemes described in subsequent Sections. However, these alternative schemes would require more parameters, and would face a more challenging optimization problem in estimating the parameters from experimental data.

The modulation scheme with sphericity has another advantage of being correlated with (bio-) physical properties, such as the amount of surface tension (stress) on the capsid, in a more intuitive way. It can potentially serve as an index, along with other proposed indices (15), to correlate physical and geometrical properties.

Deformation based on sphericity leaves the icosahedron achiral, as opposed to the scheme addressed in Sec. 3.5.3 giving rise to a chiral object.

### 3.2.2. Formulation of geometry

Consider a surface $r = r_{out}(\theta, \phi)$ that lies everywhere between the surfaces of the icosahedron $r = r_{ico}(\theta, \phi)$ and the corresponding circumscribed sphere $r = r_{sph}(\theta, \phi) = 1$. A simple case is when the distance of the new surface from the origin at each point is a linear interpolation between the distances of the icosahedral and spherical surfaces $r_{out}(\theta, \phi) = \sigma + (1 - \sigma)r_{ico}(\theta, \phi)$, where the parameter $\sigma$ is the sphericity and spans the range from 0 (icosahedron) to 1 (sphere). As shown in Sec. 3.1, $r_{ico}(\theta, \phi) = \left[\cos(\theta) + \sin(\theta)\tan(\alpha)\cos\left(\phi - \frac{\pi}{5}\right)\right]^{-1}$ over the first face of the icosahedron, and it is defined over other spherical angles $(\theta, \phi)$ using appropriate $R_{20}$ rotations. Such a purely-radial deformation leaves the range and the formulation of angular dependencies (of individual faces) intact.

With sphericity σ, the first tetrahedron will be distorted, as all linear edges (BC, BA, and CA) turn into curves, in addition to the flat cap developing a curvature. Projections of BC and BA on the x-y plane are still linear segments (as before), whereas the curved shape of CA is seen in the x-y projection.

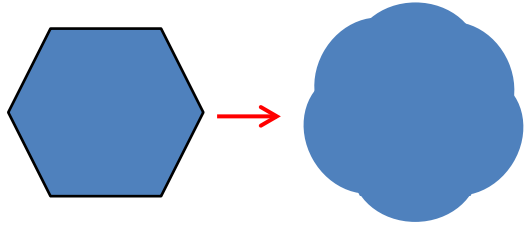

**Figure 3**: 2D projection of an icosahedron after (achiral) spherical deformation

*3.2.3. Evaluation of Fourier transform*
The Fourier transform of the deformed icosahedron $F_{S-Def-IH}(\boldsymbol{q})$ can be written similar to that of the icosahedron

$$F_{S-IH}(\boldsymbol{q}) = \int_{\phi=0}^{\phi=2\pi} d\phi \int_{\theta=0}^{\theta=\pi} \sin(\theta)\, d\theta \int_{r=0}^{r=r_{ico}(\theta,\phi)} dr\, r^2 e^{-i2\pi r q \cos(\gamma)}$$

$$F_{S-Def-IH}(\boldsymbol{q}) = \int_{\phi=0}^{\phi=2\pi} d\phi \int_{\theta=0}^{\theta=\pi} \sin(\theta)\, d\theta \int_{r=0}^{r=(1-\sigma)r_{ico}(\theta,\phi)+\sigma} dr\, r^2 e^{-i2\pi r q \cos(\gamma)}$$

Here $\gamma$ is the geodesic distance between the unit vectors $\vec{r}/r$ and $\vec{q}/q$ considered as two points on the unit sphere, with $\cos(\gamma) = \cos(\theta)\cos(\theta_q) + \sin(\theta)\sin(\theta_q)\cos(\phi - \phi_q)$. The first integration (over the radial coordinate) can be done analytically (Section X of *Supplementary Materials*). Apparent similarity with the case of unperturbed icosahedron does not facilitate angular integrations (expressing $F_{S-Def-IH}$ in terms of $F_{S-IH}$). The main challenge originates from a *nonlinear* perturbation of $r_{ico}$ as $(1-\sigma)r_{ico} + \sigma$. Derivations, details, and further discussions can be found in (Section XI of *Supplementary Materials*). The final result for the *real-part of the first tetrahedron* (and using the constants $I$ and $\Phi$ defined in Section XI of *Supplementary Materials*) is as follows

$$F_{T1/20-S-Def-IH}^{Real}(\boldsymbol{q}) = \sum_{n=0}^{\infty}(-q^2)^n \sum_{l=0}^{2n} \sin^l(\theta_q) \cos^{2n-l}(\theta_q) \sum_{m=0}^{l} I_{2n,l,m} \sin(m\phi_q + \Phi_{2n,l,m})$$

Parameters such as $I$ and $\Phi$ used in these formulations should be calculated numerically. However, such numerical calculations are one-time-only data-independent computations and can be done with arbitrarily high accuracy. They are also independent of the discrete grid of $\boldsymbol{q}$, and can be used to calculate an analytic function of any $\boldsymbol{q}$ with no need to interpolation or uniform sampling. Numerical calculation of $I$ and $\Phi$ concerns double integrals on a sphere, with the first integration (over the polar coordinate $\theta$) formulated analytically; leaving just 1 numerical integration (over the azimuthal coordinate $\phi$).

In Cartesian coordinate, even finding the boundaries of integration for 3D Fourier transform (writing integrals, not to mention evaluating them) is challenging. The simple spherical coordinate formula for the boundaries are equivalent to implicit equations for expressing $z$ in terms of $x$ and $y$, and then $x$ in terms of $y$. It results in a $12^{th}$ order polynomial equation. With change of variables and using chain's rule for double integrals, the equation is reduced to a $6^{th}$ order one, which has in principle known solutions in terms of Kampé de Fériet functions (16). Nevertheless, this is just the beginning and after all these and one analytical integration, two integrals remain to be evaluated numerically.

### 3.3. Multi-layer icosahedral shell

An icosahedral shell is the set difference between two solid icosahedrons. If the radii of the circumscribed spheres of these two spheres are denoted by $R_1$ and $R_2$ and the uniform density of the icosahedral shell by $\rho$, the real-space and Fourier-space density functions can be written as

$$f_{R_1,R_2}(\boldsymbol{r}) = \rho\big[f_{S\_IH}(\boldsymbol{r}/R_2) - f_{S\_IH}(\boldsymbol{r}/R_1)\big]$$

$$F_{R_1,R_2}(\boldsymbol{q}) = \rho\big[R_2^3 F_{S\_IH}(R_2\boldsymbol{q}) - R_1^3 F_{S\_IH}(R_1\boldsymbol{q})\big]$$

Note that as $R$ approaches zero, $F_{S\_IH}(R\boldsymbol{q})$ approaches $1/(Rq)^2$, and $R^3 F_{S\_IH}(R\boldsymbol{q})$ simply vanishes, as expected.

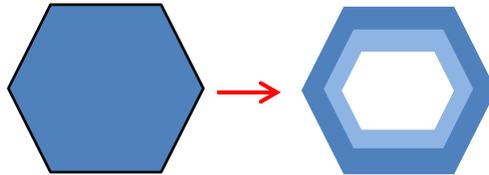

**Figure 4**: Deformation of an icosahedron towards a 2-layer empty shell

In the case of a multi-layer shell, if $R_1$ and $\rho_1$ denote the radius and the density of the innermost *solid* icosahedron and $R_n$ and $\rho_n$ the *outer* radius and density of the shell number $n$ ($2 \leq n \leq N$), one can write

$$f_{SH\_IH}(\boldsymbol{r}) = \rho_1 f_{S\_IH}(\boldsymbol{r}/R_1) + \sum_{n=2}^{N} \rho_n\big[f_{S\_IH}(\boldsymbol{r}/R_n) - f_{S\_IH}(\boldsymbol{r}/R_{n-1})\big]$$

$$F_{SH\_IH}(\boldsymbol{q}) = \rho_1 R_1^3 F_{S\_IH}(R_1\boldsymbol{q}) + \sum_{n=2}^{N} \rho_n\big[R_n^3 F_{S\_IH}(R_n\boldsymbol{q}) - R_{n-1}^3 F_{S\_IH}(R_{n-1}\boldsymbol{q})\big]$$

### 3.4. Modulated icosahedral shell
We consider two modulation schemes with sinusoidal and (orthogonal) polynomial functions.

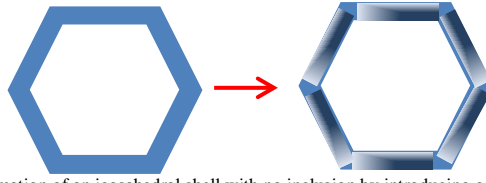

**Figure 5**: Deformation of an icosahedral shell with no inclusion by introducing gradients in the shell

#### 3.4.1. Exponential and graded multi-layer (T4-like) modulations

$$f_{Mod\_sin\_R_1,R_2}(\mathbf{r}) = f_{R_1,R_2}(\mathbf{r}) \sum_n a_n e^{i2\pi \mathbf{k}_n \cdot \mathbf{r}}$$

$$F_{Mod\_sin\_R_1,R_2}(\mathbf{q}) = \sum_n a_n F_{R_1,R_2}(\mathbf{q} - \mathbf{k}_n)$$

Sufficient condition for preserving the icosahedral symmetry is 60-fold icosahedral averaging the above expression

$$f_{Mod\_sin\_R_1,R_2\_IH}(\mathbf{r}) = f_{R_1,R_2}(\mathbf{r}) \sum_n a_n \sum_{m=1}^{60} e^{i2\pi(R_m^{-1}\mathbf{k}_n) \cdot \mathbf{r}}$$

$$F_{Mod\_sin\_R_1,R_2\_IH}(\mathbf{q}) = \sum_n a_n \sum_{m=1}^{60} F_{R_1,R_2}(\mathbf{q} - R_m^{-1}\mathbf{k}_n)$$

where $R_m$ matrices denote the 60-fold icosahedral rotation matrices (and a factor of 1/60 has been absorbed in the choice of the coefficients $a_n$). The *magnitudes* of the spatial frequencies $\mathbf{k}_n$ are determined by the relevant rates of modulations of the density function. The *directions* $\hat{\mathbf{k}}_n = \mathbf{k}_n/|\mathbf{k}_n|$, however, can be limited to the irreducible icosahedral rotation zone to remove degeneracies.

A special case of sinusoidal (complex exponential) modulations is when the faces of the icosahedron are perpendicular to the $\mathbf{k}_n$ vector(s). Such patterns are seen, for instance, in the (prolate) icosahedral capsid of $T_4$ (17). A shell with such a modulation can also be considered as a multi-layer shell with reduced number of parameters and with graded (rather than steep) density fluctuations. Such modulations are simpler to implement with individual tetrahedra forming the icosahedron:

$$f_{Mod\_sin\_R_1,R_2\_T_1/20}(\mathbf{r}) = \left(f_{T_1/20}(\mathbf{r}/R_2) - f_{T_1/20}(\mathbf{r}/R_1)\right) \sum_n a_n e^{i2\pi k_n \hat{f} \cdot \mathbf{r}}$$

$$F_{Mod\_sin\_R_1,R_2\_T_1/20}(\mathbf{q}) = \sum_n a_n \left[R_2^3 F_{T_1/20}\left(R_2(\mathbf{q} - k_n \hat{f})\right) - R_1^3 F_{T_1/20}\left(R_1(\mathbf{q} - k_n \hat{f})\right)\right]$$

Where $\hat{f} = [\sin(\alpha)\cos(\Omega/2), \sin(\alpha)\sin(\Omega/2), \cos(\alpha)]$ is the constant unit vector perpendicular to the surface of the first face of the icosahedron. The function $f_{T_1/20}(\mathbf{r})$ and hence $f_{Mod\_sin\_R_1,R_2\_T_1/20}(\mathbf{r})$ are limited to one face of the icosahedron, only. With the directions of the modulating sinusoidals taken care of, one only needs to choose appropriate scalars $k_n$; i.e., a simple *scalar* Fourier series, but without the constraint of commensurable (harmonic) values of $k_n$. For a real-valued density function, the expression is further subject to Hermitian symmetry:

$$\sum_n [a_n e^{i2\pi k_n \hat{f} \cdot \mathbf{r}} + a_n^* e^{-i2\pi k_n \hat{f} \cdot \mathbf{r}}] = \sum_n [b_n \cos(2\pi k_n \hat{f} \cdot \mathbf{r}) + d_n \sin(2\pi k_n \hat{f} \cdot \mathbf{r})]$$

A many-parameter fit using arbitrary $\mathbf{k}_n$ directions would use some unnecessary parameters to find the direction $\hat{f}$, which could have been calculated analytically beforehand. Imposing the constraint $\hat{k}_n = \hat{f}$ has effectively reduced a 3D problem to a 1D one. In fitting experimental data, this reduction has a dramatic impact on the number of parameters required to model fluctuations of the shell density. Assuming $125 = 5^3$ parameters in the general 3D formulation (without symmetry) using unknown $\mathbf{k}_n$ directions, one would need only 5 parameters to reproduce fluctuations *with the same resolution* in the known direction.

#### 3.4.2. Polynomial modulations
The Fourier transform of a function $f(\mathbf{r})$ after modulation with a 3D polynomial can be written using multivariate derivatives of its Fourier transform

$$\mathbb{F}\left\{f(\mathbf{r}) \sum_{m,n,p} A_{m,n,p} x^m y^n z^p\right\} = \sum_{m,n,p} \frac{A_{m,n,p}}{(-i2\pi)^{m+n+p}} \frac{\partial^{m+n+p}}{\partial q_x^m \partial q_y^n \partial q_z^p} \mathbb{F}\{f(\mathbf{r})\}$$

An important advantage of an analytical expression for Fourier transform is to perform such differentiations *analytically* to avoid noise-enhancement associated with numerical differentiation. Programs such as the Symbolic toolbox of Matlab can be used to perform such (straightforward, yet tedious) *analytical* differentiations.

Similar to the case of sinusoidal modulations, one can perform 60-fold averaging to maintain icosahedral symmetry. Removing degeneracies can be done by rewriting the polynomial as $\sum_{m,n,p} A_{m,n,p} x^m y^n z^p = \sum_{(m,n,p)} \mathbf{a}_{mnp} \cdot \mathbf{r}_{mnp}$, where $\mathbf{r}_{mnp}$ is a triplet of homogeneous polynomials $\mathbf{r}_{mnp} = (x^m y^n z^p, x^p y^m z^n, x^n y^p z^m)$ formed with cyclic permutation of variables, and $\mathbf{a}$ is a weight factor in $\mathbb{R}^3$. Limiting the directions of these weight factors $\hat{\mathbf{a}} = \mathbf{a}/|\mathbf{a}|$ to the irreducible rotation zone eliminates the icosahedral degeneracy (after averaging) while scanning the full $\mathbf{a}$ space ($\mathbb{R}^3$).

### 3.5. Tiled icosahedrons
The surface of an icosahedrally-symmetric convex polyhedron can be tiled (completely and without overlaps) with pentamers and hexamers of a given monomer (18). Such a tiling scheme has been used to explain the configuration of a protein array on the capsid of a virus (19).

#### 3.5.1. Quasi-equivalence
In the quasi-equivalence model, all hexamers and pentamers are initially defined on a 2D hexagonal lattice of hexamers. By choosing a specific point on the grid as the origin $O$ and traversing $h$ and then $k$ steps along two independent directions of the lattice towards a point $P$, a direction $OP$ is specified. The line segment $OP$ is one edge of the (Goldberg polyhedron, for example an) icosahedron that can be derived from the 2D hexagonal lattice.

The transition from a planar grid to a closed 3D object is done by cutting out 1/6$^{th}$ of the surface around each vertex (such as $O$ or $P$), and closing this gap by folding the grid in 3D. This procedure leaves hexamers everywhere except at the 12 vertices, where the hexamers are changed to 12 pentamers. The number of hexamers is $10(T - 1)$, where $T = h^2 + hk + k^2$ (known as the triangulation number or T-number) is a measure of the surface area and hence the size of a virus (20,21).

#### 3.5.2. Pseudo-equivalence
In Klug's *quasi*-equivalence model, the monomers are all identical. In a *pseudo*-equivalence model, a similar tiling but with different monomers is used, and a 3-fold symmetry axis of *cells* is referred to as the pseudo-3-fold axis of *monomers* (20).

#### 3.5.3. Chirality
When the characteristic indices of the hexagonal lattice are both equal or one is zero; i.e., if $hk(h - k) = 0$, the resulting icosahedron made out of the hexagonal grid is achiral, and the edges of the subunits align with the edges of the icosahedron (cases of $T = 3, 16, 25 \ldots$).

When $h$ and $k$ are both nonzero and different from each other, the resulting polyhedron can be considered as an *icosahedron with a chiral deformation of surface* (cases of $T = 7, 13, 21 \ldots$). For each T-number, there are two possible choices of chirality; denoted by subscripts $d$ and $l$ corresponding to $h < k$ and $h > k$, respectively (20).

#### 3.5.4. Symmetrons: Tiling scheme of big viruses
In 1969, Wrigley reported experimental observation and mathematical modeling of a somehow different tiling scheme with clusters in the form of deformed triangles and pentagons, which he named penta-symmetrons and tri-symmetrons, respectively (22). Later studies have categorized such architectures into three classes with specific constraints (23). Reconstructed big viruses such as PpV01, PBCV-1, and CIV are *chiral with a symmetron architecture*.

#### 3.5.5. Density of a tiled capsid in real- and Fourier-space
Each capsid protein is characterized with the triplet of $\{P_i, \boldsymbol{r}_i, R_i\}$ denoting protein type, translation vector, and rotation matrix (*preceding* the translation), respectively. Denoting the real- and Fourier- space density profiles of individual proteins (at reference orientation and centered at origin) as $f_{P_i}(\boldsymbol{r})$ and $F_{P_i}(\boldsymbol{q})$, and those of the capsid as $f(\boldsymbol{r})$ and $F(\boldsymbol{q})$, we will have

$$f(\boldsymbol{r}) = \sum_{i=1}^{N} \mathbb{T}_{\boldsymbol{r}_i} \mathbb{R}_{R_i} \{f_{P_i}(\boldsymbol{r})\} = \sum_{i=1}^{N} f_{P_i}\left(R_i^{-1}(\boldsymbol{r} - \boldsymbol{r}_i)\right)$$

$$F(\boldsymbol{q}) = \mathbb{F} \sum_{i=1}^{N} \mathbb{T}_{\boldsymbol{r}_i} \mathbb{R}_{R_i} \{f_{P_i}(\boldsymbol{r})\} = \sum_{i=1}^{N} e^{-i2\pi \boldsymbol{q} \cdot \boldsymbol{r}_i} F_{P_i}(R_i^{-1} \boldsymbol{q})$$

where the operators $\mathbb{T}$, $\mathbb{R}$, and $\mathbb{F}$ denote 3D translation, rotation, and Fourier transform, respectively, and the index $i$ runs over all capsid proteins.

At low resolutions (almost any experiment with soft X-ray energies), each capsid protein is simply a source point; i.e., $f_P(\boldsymbol{r}) = \delta(\boldsymbol{r})$ and $F_P(\boldsymbol{q}) = 1$. At such low resolutions, even a repeated building block on the capsid (trimmer made up of three proteins) can be considered as a source point.

At higher resolutions, a simple spherical, ellipsoidal, or a Gaussian profile may be more than enough. Such models may also be more efficient (at high resolutions) when used for the *asymmetric unit* of a known protein, with explicit consideration of the positions/orientations of such asymmetric units. Despite differences in amino acid sequences, many *major capsid proteins* have a double-jelly roll structure with a known translation/rotation of asymmetric units (1,24).

By approximating the *physical* information ($f_{P_i}$) with such models, the problem is reduced to a *geometric* one; i.e., the arrangement $\{\boldsymbol{r}_i, R_i\}$ of capsid proteins. In general, a position vector $\boldsymbol{r}_i$ and a rotation matrix $R_i$ each represent three nontrivial parameters, and hence a total of six:

- $R_i$: In typical cases (reconstructed capsids), the building blocks of pentasymmetrons or trisymmetrons have similar orientations, within a nontrivial rotation around the local surface normal ("in-plane" rotation considering a *locally*-flat capsid). It means only a single non-trivial parameter for $R_i$. An "in-plane" rotation scenario, as deduced from the electron microscopy reconstruction of PBCV-1, has been demonstrated in Figure 2 in (4) and Figure 2.i in (25). Such rotation scenarios of trimeric building blocks are the subject of ongoing investigation (Chuan Xiao, *Univ. of Texas at Elpaso*, personal communication). However, in low-resolution imaging with soft X-ray photons, the trimeric building blocks are simply istotropic, and there is no nontrivial parameter in $R_i$.
- $\boldsymbol{r}_i$: If the surface of the capsid is known to be a perfect icosahedron (with flat faces), the mapping from the 2D hexagonal lattice to the capsid surface and hence $\boldsymbol{r}_i$ is trivial (21-23). However, in general, even at low-resolutions, the capsid surface can be deformed. The *collective* information of all $\{\boldsymbol{r}_i\}$ corresponds to a non-trivial surface function $r_{surf}(\theta,\phi)$. Assuming icosahedral symmetry and small deviation from a perfect icosahedron, the surface function can be written as $r_{surf}(\theta,\phi) = r_{ico}(\theta,\phi) + \delta r(\theta,\phi)$, where $\delta r(\theta,\phi)$ may be approximated with a quadratic multivariate expansion with a few nontrivial parameters. Explicit formulation of scattering based on such morphological information has been addressed in (6).

### 3.6. Capsid vs. "shell"

The icosahedral symmetry of a virus originates from not only the arrangement of capsid proteins, but also from other layers (such as a lipid bilayer) underneath. With phenomenological models (such as a modulated shell), one does not have to differentiate between the capsid and such icosahedrally-symmetric layers, and they cannot be easily disentangled at the end, either. Whether a shell-model or a hybrid model (tiled capsid + lipid bilayer) is more suited for data analysis remains to be seen with specific viruses and specific achievable resolutions.

### 3.7. Icosahedron with an inclusion

The inclusion is primarily the dense (phosphor-rich) genome of the virus embedded in a solvent, as depicted in Figure 6.

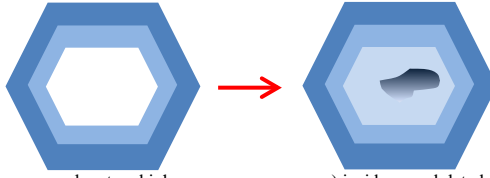

**Figure 6**: Introduction of an inclusion (homogeneous solvent and inhomogeneous genome) inside a modulated icosahedral shell (capsid, lipid bilayer, …)

#### 3.7.1. Formulation

Consider an icosahedral shell with real- and Fourier space density profiles $f_{sh}(r)$ and $F_{sh}(q)$ embedded in an otherwise-homogeneous medium with average density $\rho_0$. The inner and outer radii of the icosahedral shell are assumed to be $R_1$ and $R_2$, respectively.

Let's also assume that the inclusion is in the form of an object with the occupancy profile $o_g(r)$, embedded in a solvent with average density of $\rho_c$. $o_g(r)$ is a binary function equal to one only within genome and zero elsewhere. Let's further assume that the density function within the genome is $\rho_g[1 + m_g(r)]$; corresponding to homogeneous and non-uniform components. The total density function (by considering the *displaced* solvent, yet discarding a *uniform* background solvent) can be written as

$$f(r) = -\rho_0 f_{S-IH}(r/R_2) + f_{sh}(r)[f_{S-IH}(r/R_2) - f_{S-IH}(r/R_1)] + \rho_c[f_{S-IH}(r/R_1) - o_g(r)] + \rho_g[1 + m_g(r)]o_g(r)$$

With products giving rise to convolutions in the Fourier domain

$$F(q) = -\rho_0 R_2^3 F_{S-IH}(R_2 q) + F_{sh}(q) * [R_2^3 F_{S-IH}(R_2 q) - R_1^3 F_{S-IH}(R_1 q)] + \rho_c[R_1^3 F_{S-IH}(R_1 q) - O_g(q)] + \rho_g\{O_g(q) + M_g(q) * O_g(q)\}$$

In the simplest (and in some cases the most reliable) reconstruction, the packed genome occupies the entire icosahedral core, and all regions are homogeneous: $o_g(r) = f_{S-IH}(r/R_1)$, $m_g(r) = 0$, and $f_{sh}(r) = \rho_{shell}$ (7).

While the occupancy profile $o_g(r)$ is a full 3D *volume* density function, its nontrivial information content is that of the surface; i.e., a 2D problem. Modeling it with a resolution of $N$ points per coordinate requires $O(N^2)$ and not $O(N^3)$ unknowns.

Sinusoidal and polynomial modulations can be performed similar to the case of capsid modulation by shifting or differentiating the Fourier transform of the envelope function. However, if the genome envelope is chosen to be a 3D Gaussian, there is a simpler alternative for introducing modulations. Any polynomial of degree $N$ can be expressed in terms of orthogonal polynomials of order 0 to $N$. Among orthogonal polynomials, Hermite polynomials have the property of simple and similar modulation effects on a Gaussian in real- and Fourier spaces:

$$f_g(r) = e^{-\frac{r^2}{2R^2}} \sum_{m,n,p} a_{mnp} H_m(x/R) H_n(y/R) H_p(z/R)$$

$$F_g(q) = (R\sqrt{2\pi})^3 e^{-\frac{(2\pi R q)^2}{2}} \sum_{m,n,p} (-i)^{m+n+p} a_{mnp} H_m(2\pi R q_x) H_n(2\pi R q_y) H_p(2\pi R q_z)$$

$H_n(x) = (-1)^n e^{x^2} \frac{d^n}{dx^n} e^{-x^2}$ is the Hermite polynomial of order $n$ with "physicist's convention".

#### 3.7.2. Biological considerations: Reproducibility and "periodicity"

Relative motion or rotation of genome with respect to capsid or the deformation of genome in big viruses makes 3D reconstructions nearly impossible. On the one hand, larger viruses have more scattering atoms and more likely to generate intense images. On the other hand, among large ds-DNA viruses, the volume density of DNA base pairs decreases with size (26), which implies decreased reproducibility.

The small PBCV-1 is more likely to have a larger effective $\rho_g$ and a more secured genome (reproducible snapshots with no relative movement of capsid and genome) compared to CroV. Mimivirus has the least average density and the least expected reproducibility of the shape of the genome, despite the benefit of intense scattering.

PBCV-1 is known to have not only a ds-DNA genome, but also some proteins inside the capsid. Furthermore, some different experiments and bioinformatic analyses suggest that (ejected) viral DNA is associated with these proteins in a "periodic" fashion (26). Such an association of protein-DNA observed in the unfolded state may correspond to rotational (quasi-) symmetry of the native state. This hypothesis can be tested by applying rotational averages, as in Sec. 3.4, to $m_g(r)$.

### 3.8. Clusters of icosahedrons

Some scattering patterns from big viruses correspond to multiple viruses intercepted simultaneously by the X-ray laser pulse. A special case of such scattering scenarios corresponds to a *cluster* of icosahedral objects. Fitting experimental scattering patterns to a cluster model includes not only a continuous orientation recovery problem, but also a nontrivial constrained discrete optimization to define the cluster.

If $f_0(r)$ and $F_0(q)$ denote the real- and Fourier-space density profiles of individual identical objects (say, icosahedral shells) and $f(r)$ and $F(q)$ denote those of the cluster, one can write

$$f(r) = \sum_{i=1}^{N} \mathbb{T}_{r_i} \mathbb{R}_{R_i} \{f_0(r)\} = \sum_{i=1}^{N} f_0(R^{-1}(r - r_i))$$

$$F(q) = \mathbb{F} \sum_{i=1}^{N} \mathbb{T}_{r_i} \mathbb{R}_{R_i} \{f_0(r)\} = \sum_{i=1}^{N} e^{-i2\pi q \cdot r_i} F_0(R_i^{-1} q)$$

The problem is simply reduced to a geometrical one to find the translations and rotations of individual objects forming the cluster $\{r_i, R_i\}$. Different types of clusters formed with (interpenetrating or non-interpenetrating) icosahedra have been studied extensively in the context of alloys and Boron chemistry, with examples being the $I_3$, $L_{13}$, and $L_4$ units (27,28).

## 4. *Spherical harmonic spectrum*

An insightful representation and also way of analysis of an object is monitoring the 3D volume density (or its Fourier transform) on concentric spherical shells. In the case of an icosahedrally-symmetric object, the resulting spherical patterns show 60-fold icosahedral symmetry, as shown in Figure 7. Such icosahedrally-symmetric spherical sections through the density profile of the Sputnik virus, for instance, have been shown in Figure 2 in (29). Similar patterns can also be seen in the case of a simple homogeneous icosahedron (solid or shell). The aim of this Section is to characterize a solid icosahedron using the *spherical harmonics* basis on such spherical shells, from which both spherical and Fourier harmonics of modulated shells can be calculated.

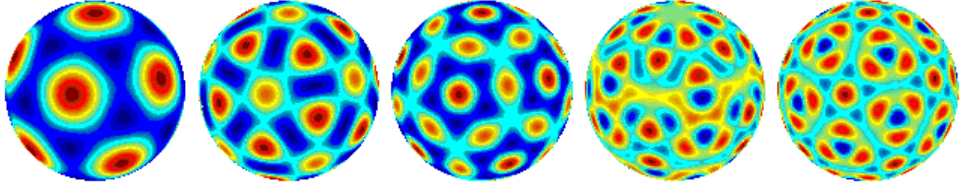

**Figure 7**: The first five nontrivial icosahedral harmonics corresponding to the orders $l \in \{6, 10, 12, 15, 18\}$.

### 4.1. Fourier and spherical spectra

At the core of Fourier transform are sinusoidal (complex exponential) functions. They correspond to the eigenfunctions of the wave (Helmholtz) equation $\nabla^2 \psi + k^2 \psi = 0$ in the Cartesian coordinate. An eigenfunction of the same equation, when formulated in the spherical coordinate, is one with separable radial and angular dependencies. The angular dependencies are described by 2-variable and 2-parameter basis of spherical harmonics $Y_{l,m}(\theta, \phi)$. A function of real-space coordinates and its 3D Fourier transform can both be decomposed into correlated spherical harmonic spectra (30).

Modeling the pattern on a spherical shell with spherical harmonics is indeed extending the JPEG model to the surface of a sphere (JPEG coding is simply projecting a BMP image on 2D cosine functions $\cos(mx)\cos(ny)$). Furthermore, these spherical harmonic representations can be converted to each other using a much simpler 1D transform; i.e., spherical Bessel (or spherical Hankel) transform. These interrelations between different transforms have been sketched in Figure 8.

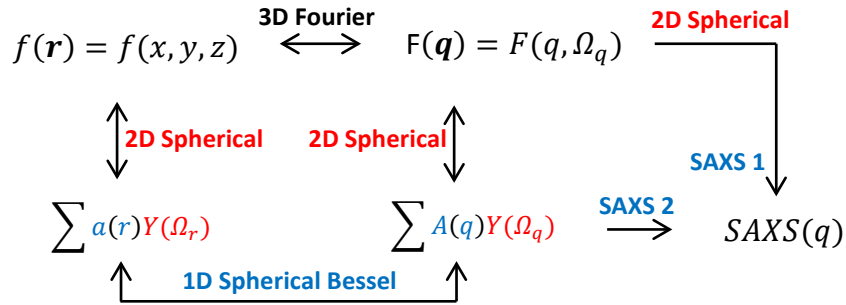

**Figure 8**: The relations between Fourier and spherical harmonics in real and reciprocal spaces, and two different ways of calculating a solution scattering (SAXS) profile (For simplicity, the summation parameters $l$ and $m$ defining a spherical harmonic order have not been shown explicitly. Furthermore, a pre-factor of $4\pi(-i)^l$ has been absorbed in the specific definition of $A(q)$).

The correlation between Fourier and spherical harmonics can be seen in the formulation of small-angle X-ray scattering (SAXS) profile. Assuming the common model of light scattering, the SAXS profile (from *solution scattering* or *powder pattern* experiments) can be obtained by either averaging the magnitude squared of 3D Fourier transform on spherical shells, or alternatively adding up the fractional powers in the spherical harmonic representation, as depicted in Figure 8. In the spherical harmonic approach, one can use a simple combination scheme, whereas in the direct approach, the synergetic term requires new nontrivial calculations (the double integral in the equation below).

$$SAXS\left\{\sum_{l,m} \alpha_{lm} Y_{lm}(\Omega_q) + \sum_{l,m} \beta_{lm} Y_{lm}(\Omega_q)\right\} = \sum_{l,m} |\alpha_{lm} + \beta_{lm}|^2$$

$$SAXS\{F_1 + F_2\} = SAXS\{F_1\} + SAXS\{F_2\} + 2 \oiint_{S^2(q)} Re\{F_1 F_2^*\} d\Omega_q$$

### 4.2. Symmetry-adapted harmonics

A spherical harmonic of order $(l_0, m_0)$ is transformed to a linear combination of spherical harmonics of orders $(l_0, \{m\})$ after an arbitrary 3D rotation (31). As such, a function expressed in terms of spherical harmonics will have a second *equivalent* spherical harmonics expansion after a symmetry-preserving rotation. Term-by-term matching of coefficients of these two expansions results in constraints on the coefficients of the spherical harmonic expansion. In the case of icosahedral symmetry, these constraints or *selection rules* can be summarized as follows (9-11):

1. The spherical harmonics $Y_{l,m}$ are allowed only in two categories of $l = 6p + 10q$ and $l = 6p + 10q + 15$, for nonzero integers $p$ and $q$; i.e., $l_{even} = 0,6,10,12,16,18,20,22,24,26,28,30, ...$ and $l_{odd} = 15,21,25,27,31, ....$ Other values of $l$ are symmetry-forbidden.
2. For a given allowed order $l$, only specific combinations $J_l = \sum_{m=-l}^{l} b_m Y_{lm}$, known as *icosahedral harmonics* or symmetry-adapted spherical harmonics, are allowed. The $b_m$ coefficients have known tabulated values (within a normalization factor), and are derived from trigonometric functions of the angle $\Omega = 2\pi/5$.
3. All spherical harmonics of a given symmetry-allowed order $l_0 < 30$ are squeezed into a *single* icosahedral harmonic $J_l(\Omega) = \sum_{m=-l}^{l} b_m Y_{lm}(\Omega)$. For $l_0 \geq 30$, there are two or more (degenerate) icosahedral harmonics as $J_{l,n}(\Omega) = \sum_{m=-l}^{l} b_{m,n} Y_{lm}(\Omega)$.

Spherical harmonic expansion of a spherical pattern can take the three following forms corresponding to no constraint, icosahedral symmetry, and low-resolution icosahedral symmetry, respectively:

$$\sum_{l=0}^{\infty} \sum_{m=-l}^{l} d_{lm} Y_{lm} \quad or \quad \sum_{l=0}^{\infty} \sum_{\mu=1}^{N_l} d_{l\mu} J_{l\mu} \quad or \quad \sum_{l=0}^{30} d_l J_l$$

Since we are concerned with low-resolution few-parameter models of an icosahedral shell, we restrict ourselves to the common assumption of limiting $l$ to 30. We note however, that the icosahedron itself has also higher orders of $l$ when formulated analytically or defined numerically on a dense grid. This is important to note with nonlinear operators (such as SAXS), as they mix different orders.

It is helpful to clarify beforehand whether the sought "density" is real or complex. While the physical *origin* of the sought density map is electron density, the very measured quantity can be different and can also be complex. In electron microscopy, the sought density map is electrostatic potential associated with electron density. In X-ray scattering at soft X-ray energies, the density map is the complex refractive index (6). In modeling a purely-real function (32), Hermitian symmetry (10) can be used to restrict orders only to the $l_{even}$ subset of selection-allowed orders.

### 4.3. Extension to full 3D basis functions

In addition to angular variations modeled with spherical harmonics (w/wo symmetry constraint), 1D functions of the radial coordinate are also required for full 3D modeling. A useful radial function is one that along with spherical harmonics makes the 3D Zernike polynomials. These functions are explicitly limited to the interior of the unit sphere in real-space and feature analytical expressions in real- and Fourier-spaces (33).

The general expansion (33) can be applied to the core and the shell of a virus separately. It can be further simplified if the inclusion density can be approximated with only radial gradients, or with angular modulations that can be approximated with icosahedral harmonics. In this case (considered in the follow-up report in the analysis of the SXAS pattern from PBCV-1), the Fourier-space profile can be expressed with icosahedral harmonics as:

$$F_{virus}(\mathbf{q}) = \sum_{n=0}^{30} \sum_{\substack{l=0 \\ n-l \text{ even} \\ l \text{ meeting selection rule}}}^{n} [\boldsymbol{\alpha}_{nl} R_{core}^3 b_n(2\pi q R_{core}) + \boldsymbol{\beta}_{nl} R_{shell}^3 b_n(2\pi q R_{shell})] J_l(\Omega_q)$$

where the function $b_n(x) = [j_n(x) + j_{n+2}(x)]/(2n + 3)$ is defined as the scaled sum of two spherical Bessel functions. Writing the SAXS profile will be facilitated by changing the order of the two summations

$$SAXS_{virus}(q) = \sum_{\substack{l=0 \\ l \text{ meeting selection rule}}}^{30} \left| \sum_{\substack{n=l \\ n-l \text{ even}}}^{30} [\boldsymbol{\alpha}_{nl} R_{core}^3 b_n(2\pi q R_{core}) + \boldsymbol{\beta}_{nl} R_{shell}^3 b_n(2\pi q R_{shell})] \right|^2$$

The number of basis functions required to model a 3D function with a "resolution" of $n_{Max} = 30$ is reduced from ~5400 (33) to ~100, when icosahedral symmetry is imposed. This number can be further reduced down to ~10, when the more complete constraint of *modulated icosahedral shell* is enforced and an efficient model of modulation/inclusion is used.

### 4.4. Icosahedral spectra

*4.4.1. Icosahedral spectrum of an icosahedron*
Having evaluated the density profiles of the unit solid icosahedron in real and Fourier spaces analytically, we can calculate its spherical spectrum on a discrete grid with arbitrarily high accuracy, as a one-time-only problem. The result is discrete set of 1D functions $a_{l,S-IH}(r)$ and $A_{l,S-IH}(q)$ profiles, as defined in Figure 8. $l_{even} = \{0,6,10,12,16,18,20,22,24,26,28,30\}$ represents the small set of 12 values that the discrete parameter $l$ can assume for a real function. Furthermore, if the radial profile is also expressed with a basis, such as those corresponding to Zernike polynomials, then the information content of $a_l(r)$ or $A_l(q)$ profiles, is simply reduced to a sparse 12 × N matrix (similar to coefficients $\beta_{nl}$ in Sec. 4.3).

*4.4.2. Icosahedral spectrum of an icosahedral shell*
With known icosahedral spectrum of a solid icosahedron ($\rho = 1$, $R_{ico} = 1$), the icosahedral spectrum of a shell with inner and outer radii $R_1$ and $R_2$ and a homogenous density $\rho$, can also be calculated

$$a_{l,S-IH}(r) = \oiint u[r_{ico}(\Omega) - r]J_l(\Omega)d\Omega$$

$$a_l(r) = \rho \oiint \{u[r_{ico}(\Omega) - r/R_2] - u[r_{ico}(\Omega) - r/R_1]\}J_l(\Omega)d\Omega$$

$$a_l(r) = \rho[a_{l,S-IH}(r/R_2) - a_{l,S-IH}(r/R_1)]$$

$$A_l(q) = \rho[R_2^3 A_{l,S-IH}(R_2 q) - R_1^3 A_{l,S-IH}(R_1 q)]$$

$a_l(r)$ has non-zero values only for $R_1 \cos(\alpha) \leq r \leq R_2$.
For a multi-layer icosahedral shell defined before, one can write:

$$f_{SH\_IH}(\mathbf{r}) = \rho_1 f_{S\_IH}(\mathbf{r}/R_1) + \sum_{n=2}^{N} \rho_n [f_{S\_IH}(\mathbf{r}/R_n) - f_{S\_IH}(\mathbf{r}/R_{n-1})]$$

$$a_l(r) = \rho_1 a_{l,S-IH}(r/R_1) + \sum_{n=2}^{N} \rho_n [a_{l,S-IH}(r/R_n) - a_{l,S-IH}(r/R_{n-1})]$$

$$A_l(q) = \rho_1 R_1^3 A_{l,S-IH}(R_1 q) + \sum_{n=2}^{N} \rho_n [R_n^3 A_{l,S-IH}(R_n q) - R_{n-1}^3 A_{l,S-IH}(R_{n-1} q)]$$

The icosahedral spectrum of an icosahedral shell with inner and outer radii of $R_1 = 0.8$ and $R_2 = 1$ has been shown in Figure 9.

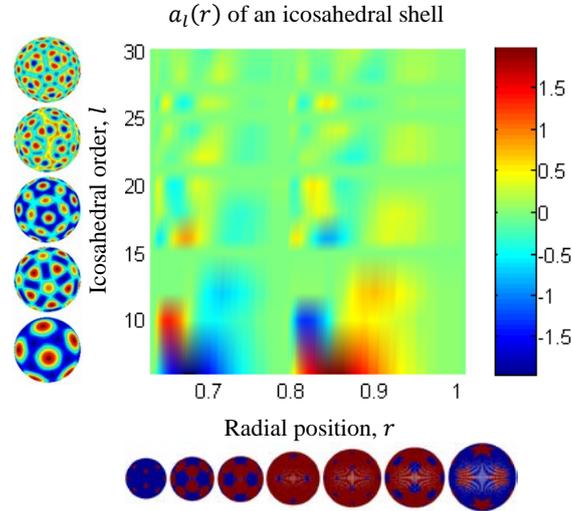

**Figure 9**: 2D representation of the full 3D information of an icosahedral shell. The horizontal axis represents the radius of a spherical shell ($R_1 = 0.8$ and $R_2 = 1$), and the vertical axis represents the icosahedral spectrum of the shell at given radius; i.e., the weights of icosahedral harmonics (non-degenerate orders $l > 0$). In the false-color 2D representation, interpolated values in-between discrete selection-allowed values of icosahedral order (for instance $6 < l < 10$) are mere guide to eye and have no quantitative meaning.

*4.4.3. Icosahedral and Fourier spectra of radially deformed icosahedra*
The icosahedral spectrum of a deformed solid icosahedron with non-zero sphericity (Sec. 2.3) can be written as

$$a_l(r) = \oiint u[\sigma + (1 - \sigma)r_{ico}(\Omega) - r]J_l(\Omega)d\Omega$$

The smallest value of $r_{ico}(\Omega)$ is $r_{min} = \cos(\alpha)$, and hence the smallest distance to the surface of the deformed icosahedron is $r_{min}^\sigma = \sigma + (1 - \sigma)\cos(\alpha)$.

For $r \leq r_{min}^\sigma$, a spherical shell with a radius $r$ lies completely within the deformed icosahedron. The argument of the step function is positive for all $\Omega$, and hence $a_l(r) = \oiint J_l(\Omega)d\Omega = \sqrt{4\pi}\delta_{l,0}$.

For $1 < r$, a spherical shell with a radius $r$ lies completely outside the deformed icosahedron. The argument of the step function is negative for all $\Omega$, and hence $a_l(r) = 0$.

For $r_{min}^\sigma < r < 1$, a spherical shell with a radius $r$ lies partially within the deformed icosahedron. The double integral is then limited to 20 similar spherical caps that lie inside the deformed icosahedron. Given the icosahedral symmetry and known analytical equation for $r_{ico}(\Omega)$ over the first tetrahedron ($T_1$), one can write

$$a_l(r) = \iint_{\{\Omega|\, r_{ico}(\Omega) > \frac{r-\sigma}{1-\sigma}\}} J_l(\Omega)d\Omega = 20 \iint_{\{\Omega|\, r_{ico}(\Omega) > \frac{r-\sigma}{1-\sigma}\ \&\ \Omega \in T_1\}} J_l(\Omega)d\Omega$$

Using the expression for $r_{ico}$ and the limits of $\Omega = (\theta, \phi)$ over $T_1$ (Section 3.1), analytical expressions for the *range* of the second double integral can be expressed as

$$0 < \phi < \Omega$$
$$0 < \theta < \theta_{CA}(\phi) \quad \& \quad \sin[\theta + Q(\phi)] < \sin[Q(\phi)] * (1-\sigma)/(r-\sigma)$$

where $Q(\phi) = \cot^{-1}[\tan(\alpha)\cos(\phi - \Omega/2)]$.

### 4.5. Linear vs. nonlinear perturbations in icosahedral spectrum

Knowing the icosahedral spectrum of an icosahedral shell (as discrete set of radial functions), one can simply add arbitrary 1D perturbations to such spectrum to tailor the 3D profile while preserving the icosahedral symmetry:

$$a_l(r) = a_{l,SH-IH}(r) + \delta_l(r)$$
$$A_l(q) = A_{l,SH-IH}(q) + \Delta_l(q)$$

Contrary to such linear order-preserving additive scheme, the product of two icosahedrally-symmetric functions is a nonlinear operation, which generates new orders as

$$J_{l1}(\Omega)J_{l2}(\Omega) = \sum_l D_{l,l_1,l_2} J_l(\Omega)$$

The derivation and the definition of expansion coefficients $D_{l,l_1,l_2}$ *with the assumption of non-degeneracy (low resolution)* have been detailed in Section XII of *Supplementary Materials*. This equation implies that starting with a pair of single lines in the icosahedral spectrum, the modulated signal comprises multiple lines (as opposed to Fourier harmonics, for which the product of $l_1$ and $l_2$ harmonics is the single $l_1 + l_2$ harmonic). In analogy with telecommunications theory, this property is similar to that of frequency-modulated (FM) and not amplitude-modulated (AM) signals (34).

In single particle imaging with electron microscopy, if the object is formulated as a linear perturbation with the maximum order $l$, the measured data is also formulated linearly and with the same maximum order. In X-ray scattering, however, the nonlinearity of loss of phase challenges both properties and hence the uniqueness of the sought 3D structure (up to a specific "resolution"). The maximum measured order $l$ corresponds to multiple orders, including missing higher-order ones (6).

### 4.6. Non-icosahedral functions for modeling distinguished capsid vertices (portals)

Minor deviation of a virus from icosahedral symmetry can imply significant biological information (distinguished portal(s) involved in and adapted to the process of infection), and yet weak signature in experimental data (potentially comparable or even smaller than noise or background level). Direct observation (Figure S1 in Supplementary Information of (35)) and 3D reconstruction with reduced symmetry constraint (35) have shown the existence of such a unique icosahedral vertex in the big virus PBCV-1.

The non-icosahedral component of a function defined on a spherical shell has the following properties in its spherical harmonic expansion (Section XIII of *Supplementary Materials*):

$$f(\Omega) = f_{ico}(\Omega) + f_{non-ico}(\Omega) = f_{ico}(\Omega) + \sum_{l=0}^{\infty}\sum_{m=-l}^{l} a_{l,m} Y_{lm}(\Omega)$$

- Different orders $l$ can be treated separately, and are not coupled.
- For symmetry-forbidden orders $l$, there is no constraint on $a_{l,m}$.
- For a non-degenerate symmetry-allowed order $l$, the vector of coefficients $\boldsymbol{a}_l$ is orthogonal to the vector of coefficients $\boldsymbol{b}_l^*$ (defined in Section 4.2); i.e., $\boldsymbol{a}_l \cdot \boldsymbol{b}_l^* = \sum_{m=-l}^{l} b_{l,m}^* a_{l,m} = \sum_{m=-l}^{l} b_{l,-m} a_{l,m} = 0$
  - Given zero values of $b_{l,m}$ for values of $m$ that are not integer multiples of 5, the constraint is further limited to $a_{l,5m}$.
- Enforcing the *new* constraint of 5-fold symmetry on a non-icosahedral function (35), requires all $a_{l,m}$ to be zero, except $a_{l,5m}$ (irrespective of whether $l$ is symmetry-allowed or symmetry-forbidden).

## 5. *Reflections and resonances of an icosahedron*

### 5.1. Electromagnetic context

The common (Geometrical) model of X-ray scattering assumes that the incident light simply sweeps the scattering object without being affected (5). The *transmitted* light (coincident with scattered light at $|\boldsymbol{q}| = 0$) is nearly the same as the incident light, and the much weaker scattered light is a minor perturbation. There are two specific cases that this assumption is challenged

- An incident plane wave is completely *reflected* at grazing incidence (incident light nearly parallel with the surface being illuminated). Total external reflection can start at angles of ~ 2 - 5 degrees, in which case even the transmitted light is zero. At slightly larger angles, the reflection is not complete, but it can still be considerable.
- At normal incidence or large angles between the incident light and the plane of incidence, reflection coefficient is small. Even a single reflection can be discarded, and multiple reflections can be very weak. However, multiple reflections caused by the faces of a symmetric object (especially when combined with total reflection) may form a *geometrical loop* and an *electromagnetic resonance*. In excitations with a (broadband) pulse laser, even more than one such resonance frequencies can be excited. An insightful analogy for understanding resonances of an icosahedron is the relevant and simpler case of 2D hexagonal object, for which the resonances have been studied theoretically (36) and investigated experimentally, especially in the case of ZnO lasers (37).

Using ray optics (and hence the formalism developed here) for such problems is only for estimation and intuitive interpretation of the results that require electromagnetic formulation in a rigorous sense.

## 5.2. Basic geometries of scattering objects

Optical imaging of a hydrated virus can be associated with one or multiple occurrences of reflections from icosahedral, spherical, and cylindrical faces. The last two cases, corresponding to a hydration shell and a liquid jet, have been addressed in Section XIV of *Supplementary Materials*.

The goal here is to derive analytical expressions for reflections from an illuminated solid homogeneous icosahedron at a given orientation. Reflections off such an icosahedral object (solid or shell) can be further split into a series of reflections from triangular interfaces.

## 5.3. Reflection off a triangular plane interface

Consider an arbitrary triangular face with vertices $A, B, C$ and an edge length of $l$, as shown in Figure 10. An arbitrary point on such a triangle is expressed as a convex interpolation of two edges with a pair of parameters $(\lambda, \mu)$, as

$$\vec{r}_{Tri} = \overrightarrow{OM} = \overrightarrow{OA} + \overrightarrow{AM} = \overrightarrow{OA} + \lambda\overrightarrow{AB} + \mu\overrightarrow{AC},$$

where $0 \leq \lambda, \mu, 1 - (\lambda + \mu) \leq 1$.

An arbitrary ray can also be modeled as $\vec{r}_l = \vec{P}_0 + t * \hat{d}$, where $\vec{P}_0$ is an arbitrary point on the line (ray); $\hat{d}$ is a (dimensionless and unitless) unit-vector along the ray *and in the direction of light propagation*. The free non-negative parameter $t$ (with the dimension of length and the same unit as $\vec{r}_l$) specifies arbitrary points on the ray in the direction of propagation.

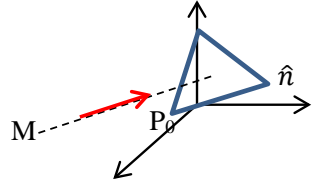

**Figure 10**: Visualization of one face of a given icosahedron and an arbitrary ray

The necessary condition for having reflection off the triangle ABC is to have a point of intersection, which should satisfy the equations of the line and the triangle simultaneously. By rewriting this vector equation as a matrix multiplication, the final results for surface and line parameters of the point of intersection are determined to be

$$\overrightarrow{OA} + \lambda\overrightarrow{AB} + \mu\overrightarrow{AC} = \vec{P}_0 + t\hat{d} \Rightarrow [\overrightarrow{AB} \quad \overrightarrow{AC} \quad -l\hat{d}]\begin{bmatrix}\lambda\\\mu\\t/l\end{bmatrix} = (\vec{P}_0 - \overrightarrow{OA}) \Rightarrow \begin{bmatrix}\lambda\\\mu\\t/l\end{bmatrix} = [\overrightarrow{AB} \quad \overrightarrow{AC} \quad -l\hat{d}]^{-1}(\vec{P}_0 - \overrightarrow{OA})$$

where each vector is considered as a 3x1 *column* matrix, and the line is assumed *not* to be coplanar with the triangle; i.e., the matrix $[\overrightarrow{AB} \quad \overrightarrow{AC} \quad -l\hat{d}]$ is not singular. The solutions should be evaluated subject to constraints $0 < \lambda, \mu, 1 - (\lambda + \mu) < 1$ and $t > 0$. Hitting an edge or a vertex corresponds to the extreme cases of these inequalities, and requires a different treatment and explicit *electromagnetic* assumptions about handling the singularity.

In case the line is coplanar with the triangle, there can be zero, one, or infinitely many solutions (corresponding to the line lying outside the triangle, hitting a vertex, or passing through the triangle).

Assuming a point of intersection on one face (ABC) of an icosahedron, the sufficient condition for having total reflection (a "streak") off the given face of the icosahedron is determined by the angle between the direction of the incoming light and the vector normal to that surface: $\theta_{grazing\ incidence} = \sin^{-1}(\hat{n}.\hat{d})$, where the *inward* surface normal vector is $\hat{n} = -(\overrightarrow{AB} \times \overrightarrow{AC})/|\overrightarrow{AB} \times \overrightarrow{AC}|$. For a given face $V_iV_jV_k$ of an icosahedron, the choice of the vertices as $A$, $B$, and $C$ is arbitrary, as long as $\overrightarrow{AB} \times \overrightarrow{AC}$ is an *outward*-pointing vector. It limits the 6 available permutations to 3 allowed ones.

Typically, total external reflection of X-rays happens at a critical angle of ~$2 - 5$ degrees, comparable with the range corresponding to the scattered component (hence the superposition of the two in measured snapshots). Knowing the directions of the incident and normal vectors, the directions of reflected and transmitted waves can be calculated using the law of reflection, as exemplified for a sphere in Section XIV of *Supplementary Materials*.

In practice, the incident light has a diameter larger than the object and is not a ray, per se. Such a beam can be split into a *bundle* of discrete parallel rays. The simple ray-tracing scheme formulated here can be applied to each ray within the bundle, with fine discretization of bundle for rays close to edges. Furthermore, the incident light can have a (Gaussian-like) distribution of intensity and even a phase profile. A nonlinear phase profile (curvature of phase front) simply means different $\hat{d}$ directions across the transverse beam profile. Aside from nonuniform intensity and phase across the incident beam, the reflection coefficient for each ray is also a complex number characterizing a gain and a phase shift (Fresnel coefficients).

When this formalism is applied to an icosahedron, a ray might *geometrically* hit points within two faces. *Electromagnetically*, however, the incident ray only hits the closer face (corresponding to a smaller value of $t$). Other faces can then be hit by the beam transmitted through the first face.

## 6. Concluding remarks

Aside from general numerical considerations regarding a model with many parameters (over-fitting, curse of dimensionality …), the problem of inverse X-ray scattering faces the uniqueness issue due to loss of phase and truncation of scattering patterns. Imposing meaningful constraints, as in the established case of protein crystallography, is not just a matter of convenience for improving signal to noise ratio. It can be a fundamental requirement to create a unique solution.

In a low-resolution 3D reconstruction, the number of required basis functions can be reduced from ~5400 to ~100 (~ 60-fold decrease), when icosahedral symmetry is enforced. This number can be further reduced down to ~10, when the more complete constraint of *modulated icosahedral shell* is enforced.

The formulations detailed in this contribution provide a *tool* for easy and controllable tradeoff between the number of parameters on the one hand and unbiased analysis on the other hand. Identification of the "optimal" tradeoff and validation of the results require further inputs and analyses.

## *Acknowledgements*

The author is grateful to Matthias Fischer and Robert L. Shoeman for frequent discussions of biological aspects, and to Bruce Doak for discussions of physical aspects.

## *References*

1. Chuan, X. and M. G. Rossmann. 2011. Structures of giant icosahedral eukaryotic dsDNA viruses. Current Opinion in Virology. 1:101-109.
2. Seibert, M. M., T. Ekeberg, F. R. N. C. Maia, M. Svenda, J. Andreasson, O. Jönsson, D. Odic, B. Iwan, A. Rocker, D. Westphal, M. Hantke, D. P. DePonte, A. Barty, J. Schulz, L. Gumprecht, N. Coppola, A. Aquila, M. Liang, T. A. White, A. Martin, C. Caleman, S. Stern, C. Abergel, V. Seltzer, J. Claverie, C. Bostedt, J. D. Bozek, S. Boutet, A. A. Miahnahri, M. Messerschmidt, J. Krzywinski, G. Williams, K. O. Hodgson, M. J. Bogan, C. Y. Hampton, R. G. Sierra, D. Starodub, I. Andersson, S. Bajt, M. Barthelmess, J. C. H. Spence, P. Fromme, U. Weierstall, R. Kirian, M. Hunter, R. B. Doak, S. Marchesini, S. P. Hau-Riege, M. Frank, R. L. Shoeman, L. Lomb, S. W. Epp, R. Hartmann, D. Rolles, A. Rudenko, C. Schmidt, L. Foucar, N. Kimmel, P. Holl, B. Rudek, B. Erk, A. Hömke, C. Reich, D. Pietschner, G. Weidenspointner, L. Strüder, G. Hauser, H. Gorke, J. Ullrich, I. Schlichting, S. Herrmann, G. Schaller, F. Schopper, H. Soltau, K. Kühnel, R. Andritschke, C. Schröter, F. Krasniqi, M. Bott, S. Schorb, D. Rupp, M. Adolph, T. Gorkhover, H. Hirsemann, G. Potdevin, H. Graafsma, B. Nilsson, H. N. Chapman, and J. HajduSeibert. 2011. Single mimivirus particles intercepted and imaged with an X-ray laser. Nature. 470:78-81.
3. Fischer, M. G., M. J. Allen, W. H. Wilson, and C. A. Suttle. 2010. Giant virus with a remarkable complement of genes infects marine zooplankton. Proc. Natl. Acad. Sci. USA. 107:19508-19513.
4. Zhang, X., Y. Xiang, D. D. Dunigan, T. Klose, P. R. Chipman, J. L. Van Etten, and M. G. Rossmann. 2011. Three-dimensional structure and function of the Paramecium bursaria chlorella virus capsid. Proc. Natl. Acad. Sci. USA. 108:14837-14842.
5. Jafarpour, A. 2014. On X-ray scattering model for single particles, Part I: The legacy of protein crystallography. arXiv:1407.6595 [Quantitative Biology.Biomolecules]. http://arxiv.org/abs/1407.6595.
6. Jafarpour, A. 2014. On X-ray scattering model for single particles, Part II: Beyond protein crystallography. arXiv:1407.6596 [Quantitative Biology.Biomolecules]. http://arxiv.org/abs/1407.6596.
7. Li, X., C.Y. Shew, L. He, F. Meilleur, D. A. A. Myles, E. Liu, Y. Zhang, G. S. Smith, K. W. Herwig, R. Pynn, and W. R. Chen. 2011. Scattering functions of Platonic solids. J. Appl. Cryst. 44:545–557.
8. Kramer, P. and D. Zeidler. 1989. Structure factors for icosahedral quasicrystals. Acta Cryst. A45:524-533.
9. Jack, A. and S. C. Harrison. 1975. On the interpretation of small-angle X-ray solution scattering from spherical viruses. J. Mol. Biol. 99:15–25.
10. Liu, H., L. Cheng, S. Zeng, C. Cai, Z. H. Zhou, and Q. Yang. 2008. Symmetry-adapted spherical harmonics method for high-resolution 3D single-particle reconstructions. J. Struct. Biol. 161:64–73.
11. Zheng, Y., P. C. Doerschuk, and J. E. Johnson. 1995. Determination of three-dimensional low-resolution viral structure from solution x-ray scattering data. Biophys. J. 69:619–639.
12. Baldwin, P. R. and P. A. Penczek. 2007. The Transform Class in SPARX and EMAN2. J. Struct. Biol. 157: 250–261.
13. Baker, T. S., N. H. Olson, and S. D. Fuller. 1999. Adding the Third Dimension to Virus Life Cycles: Three-Dimensional Reconstruction of Icosahedral Viruses from Cryo-Electron Micrographs. Microbiol. Mol. Biol. Rev. 63:862-922.
14. Chimera. *http://www.cgl.ucsf.edu/chimera/docs/ContributedSoftware/icosahedron/icosahedron.html*.

15. Lidmar, J., L. Mirny, and D. R. Nelson. 2003. Virus shapes and buckling transitions in spherical shells. Phys. Rev. E. 68:051910.
16. Weisstein, E. W. http://mathworld.wolfram.com/SexticEquation.html.
17. Klug, W. S. 2003. *A Director-Field Theory of DNA Packaging in Bacteriophage Viruses.* The Library of California Institute of Technology. http://thesis.library.caltech.edu/4059/
18. Goldberg, M. 1937. A Class of Multi-Symmetric Polyhedra. Tohoku Math. J . 43:104-108.
19. Caspar, D. L., and A. Klug. 1962. Physical principles in the construction of regular viruses. Cold Spring Harbor Symp. Quant. Biol. 27:1-24.
20. ViperdB. *http://viperdb.scripps.edu/virus.php*.
21. Mannige, R. V. and C. L. Brooks III. 2010. Periodic Table of Virus Capsids: Implications for Natural Selection and Design. PLoS ONE. 5:e9423.
22. Wrigley, N. G. 1969. An Electron Microscope Study of the Structure of Sericesthis Iridescent Virus. J. Gen. Virol. 5:123-134.
23. Sinkovits, R. S. and T. S. Baker. 2010. A tale of two symmetrons: Rules for construction of icosahedral capsids from trisymmetrons and pentasymmetrons. J. Struct. Biol. 170:109-116.
24. Krupovic, M., D. H. Bamford, and E. V. Koonin. 2014. Conservation of major and minor jelly-roll capsid proteins in Polinton (Maverick) transposons suggests that they are bona fide viruses. Biology Direct. 9:6.
25. Lawson, C. L., S. Dutta, J. D. Westbrook, K. Henrick, and H. M. Berman. 2008. Representation of viruses in the remediated PDB archive. Acta Cryst. D64:874-882.
26. Wulfmeyer, T., C. Polzer, G. Hiepler, K. Hamacher, R. Shoeman, D. D. Dunigan, J. L. Van Etten, M. Lolicato, A. Moroni, G. Thiel, and T. Meckel. 2012. Structural Organization of DNA in Chlorella Viruses. PLoS ONE. 7:e30133.
27. Makongo Mangan, J. P. 2009. A case study of complex metallic alloy phases: structure and disorder phenomena of Mg-Pd compounds. *http://d-nb.info/993377696/34*.
28. Wikipedia. http://en.wikipedia.org/wiki/Crystal_structure_of_boron-rich_metal_borides.
29. Sun, S., B. La Scola, V. D. Bowman, C. M. Ryan, J. P. Whitelegge, D. Raoult, and M. G. Rossmann. 2010. Structural Studies of the Sputnik Virophage. J. Virol. 84:894-897.
30. Baddour, N. .2010. Operational and convolution properties of three-dimensional Fourier transforms in spherical polar coordinates. J. Opt. Soc. Am. A. 27:2144-2155.
31. Wigner, E. 1931. Gruppentheorie und ihre Anwendung auf die Quantenmechanik der Atomspektren. Springer Verlag.
32. Saldin, D. K., H.-C. Poon, P. Schwander, M. Uddin, and M. Schmidt. 2011. Reconstructing an icosahedral virus from single-particle diffraction experiments.. Opt. Express. 19:17318-17335.
33. Liu, H., B. K. Poon, A. J. E. M. Janssen, and P. H. Zwart. 2012. Computation of fluctuation scattering profiles via three-dimensional Zernike polynomials. Acta Cryst. A68:561-567.
34. Carlson, A. B. 2001. Communication Systems: An introduction to signals and noise in electrical communication. McGraw-Hill.
35. Cherrier, M. A., V. A. Kostyuchenko, C. Xiao, V. D. Bowman, A. J. Battisti, X. Yan, P. R. Chipman, T. S. Baker, J. L. Van Etten, M. G. Rossmann. 2009. An icosahedral algal virus has a complex unique vertex decorated by a spike. Proc. Natl. Acad. Sci. USA. 106:11085-11089.
36. Wiersig, J. 2003. Hexagonal dielectric resonators and microcrystal lasers. Phys. Rev. A. 67:023807.
37. Wang, N., X. Chen, Y. Yang, J. Dong, C. Wang, and G. Yang. 2013. Diffuse reflection inside a hexagonal nanocavity. Scientific Reports. 3:1298.
#38. Zheng, Y. 1994. *http://docs.lib.purdue.edu/cgi/viewcontent.cgi?article=1206&context=ecetr*.
#39. Lebedev, V. I. 1977. Spherical quadrature formulas exact to orders 25–29. Siberian Math. J. 18:99-107.

## *Supplementary Information*

### I. $R_{20}$ and $R_{60}$ rotation matrices

In principle, all 60 rotation matrices of the icosahedral group can be made with combinations of Rz(Ω) and Ry(θ$_c$) (12). However, it is more intuitive to include a 3$^{rd}$ generator Ry(π) (S.38) to rotate any triangle "upside-down" directly. Note that even though the two matrices Rz(Ω) and Ry(θ$_c$) are *linearly* independent, they are correlated via three *nonlinear* equations (12).

Here, we use an intuitive approach and use explicit knowledge of icosahedron vertices to find the rotation matrices. Any rotation matrix is the solution to 3 vector equations with 9 unknowns. If $v_1$, $v_2$, and $v_3$ coincide with $v_m$, $v_n$, and $v_p$ after rotation, one should have $v_m=Rv_1$, $v_n=Rv_2$, and $v_p=Rv_3$, or by combining matrix multiplications, R[$v_1$ | $v_2$ | $v_3$]=[$v_m$ | $v_n$ | $v_p$], and finally R=[$v_m$ | $v_n$ | $v_p$][$v_1$ | $v_2$ | $v_3$]$^{-1}$.

The employed combination of the indices (m,n,p) for different faces $F_l$ (1 ≤ l ≤ 20) are as follows (permutation is irrelevant):

$F$1(1,2,3), $F$2(1,3,4), $F$3(1,4,5), $F$4(1,5,6), $F$5(1,6,2),
$F$6(2,3,10), $F$7(3,4,11), $F$8(4,5,7), $F$9(5,6,8), $F$10(6,2,9),
$F$11(3,10,11), $F$12(4,11,7), $F$13(5,7,8), $F$14(6,8,9), $F$15(2,9,10),
$F$16(7,8,12), $F$17(8,9,12), $F$18(9,10,12), $F$19(10,11,12), $F$20(11,7,12)

### II. Boundaries of the first (irregular) tetrahedron $T_{1/20}$

The *Cartesian* coordinates of the vertices of the first tetrahedron are as follows:

$$O[0,0,0], B[0,0,1], C[\sin(\theta_c),0,\cos(\theta_c)], A[\sin(\theta_c)\cos(\Omega),\sin(\theta_c)\sin(\Omega),\cos(\theta_c)]$$

Each face of the tetrahedron is characterized with 1) the equation of the *plane* on which it lies, and 2) the equation of the boundary *curve(s)*; i.e., the edges of the triangle.

*Boundaries of the first tetrahedron in the spherical coordinate*
The entire volume of the first tetrahedron is defined with the following boundaries in the *spherical* coordinate
$$0 \leq \phi \leq \Omega$$
$$0 \leq \theta(\phi) \leq \theta_{CA}(\phi)$$
$$0 \leq r(\theta,\phi) \leq r_{ABC}(\theta,\phi)$$

**The CA edge of tetrahedron $\theta_{CA}(\phi)$**
The equation of the *line segment* CA can be derived in the *Cartesian* coordinate as

$$\frac{x-\sin(\theta_c)}{\sin(\theta_c)\cos(\Omega)-\sin(\theta_c)} = \frac{y-0}{\sin(\theta_c)\sin(\Omega)-0}, z=\cos(\theta_c)$$

$$\frac{x-\sin(\theta_c)}{y} = \frac{\sin(\theta_c)\cos(\Omega)-\sin(\theta_c)}{\sin(\theta_c)\sin(\Omega)-0}, r=\cos(\theta_c)/\cos(\theta)$$

$$\frac{x-\sin(\theta_c)}{y} = \frac{-2\sin^2(\Omega/2)}{2\sin(\Omega/2)\cos(\Omega/2)} = -\tan(\Omega/2), r=\cos(\theta_c)/\cos(\theta)$$

$$\frac{r\sin(\theta)\cos(\phi)-\sin(\theta_c)}{r\sin(\theta)\sin(\phi)} = -\tan(\Omega/2), r=\cos(\theta_c)/\cos(\theta)$$

Now, by eliminating the variable $r$, the *spherical* coordinate equation of the *line segment* CA is written as
$$\frac{\cos(\phi)-\sin(\theta_c)/[\sin(\theta)\cos(\theta_c)/\cos(\theta)]}{\sin(\phi)} = -\tan(\Omega/2)$$
$$\frac{\cos(\phi)-\tan(\theta_c)\cot(\theta)}{\sin(\phi)} = -\tan(\Omega/2)$$

$$\tan(\theta_c)\cot(\theta) = \cos(\phi)+\tan(\Omega/2)\sin(\phi) = \cos(\phi-\Omega/2)/\cos(\Omega/2)$$

$$\theta = \theta_{CA}(\phi) = \cot^{-1}[\cot(\theta_c)\sec(\Omega/2)\cos(\phi-\Omega/2)]$$

**The ABC face of tetrahedron $r_{ABC}(\theta,\phi)$**
Consider an arbitrary point M on the face ABC with the *spherical* coordinate $M(r,\theta,\phi)$. Since the three vectors $\overrightarrow{BM}, \overrightarrow{BA}$, and $\overrightarrow{BC}$ are coplanar, one can write

$$\overrightarrow{BM}.(\overrightarrow{BA}\times\overrightarrow{BC}) = \begin{vmatrix} r\sin(\theta)\cos(\phi)-0 & r\sin(\theta)\sin(\phi)-0 & r\cos(\theta)-1 \\ \sin(\theta_c)\cos(\Omega)-0 & \sin(\theta_c)\sin(\Omega)-0 & \cos(\theta_c)-1 \\ \sin(\theta_c)-0 & 0-0 & \cos(\theta_c)-1 \end{vmatrix} = 0$$

$$\Rightarrow r = r_{ABC}(\theta,\phi) = [\cos(\theta)+\sin(\theta)\tan(\alpha)\cos(\phi-\Omega/2)]^{-1}$$

*Boundaries of the first tetrahedron in the Cartesian coordinate*
The entire volume of the first tetrahedron is defined with the projection of the BCA triangle on the $xy$ plane, and the $z$-span over this area:
$$0 \leq y \leq y_{A\prime}$$
$$x_{B\prime A\prime}(y) \leq x(y) \leq x_{C\prime A\prime}(y)$$
$$z_{OCA}(x,y) \leq z(x,y) \leq z_{BCA}(x,y)$$

**Boundaries of the $xy$ projection**
$x_{B\prime A\prime}(y)$ and $x_{C\prime A\prime}(y)$ can be easily found using the coordinates of B, C, A and their projections:
$O[0,0,0], B[0,0,1], C[\sin(\theta_c),0,\cos(\theta_c)], A[\sin(\theta_c)\cos(\Omega),\sin(\theta_c)\sin(\Omega),\cos(\theta_c)]$
$O\prime[0,0], B\prime[0,0], C\prime[\sin(\theta_c),0], A\prime[\sin(\theta_c)\cos(\Omega),\sin(\theta_c)\sin(\Omega)]$
So, $x_{B\prime A\prime}(y) = \cot(\Omega)\,y$ and $x_{C\prime A\prime}(y) = \sin(\theta_c) - \tan(\Omega/2)\,y$

**The two planes underneath and on top**
The equation of the plane passing through an arbitrary point Q and three fixed points M, N, and P is $\overrightarrow{MQ}.(\overrightarrow{MN}\times\overrightarrow{MP}) = 0$.
So, the equations of the two planes OCA and BCA is

OCA: $\overrightarrow{OM}.(\overrightarrow{OC}\times\overrightarrow{OA}) = \begin{vmatrix} x & y & z \\ \sin(\theta_c) & 0 & \cos(\theta_c) \\ \sin(\theta_c)\cos(\Omega) & \sin(\theta_c)\sin(\Omega) & \cos(\theta_c) \end{vmatrix} = 0$

BCA: $\overrightarrow{BM} \cdot (\overrightarrow{BC} \times \overrightarrow{BA}) = \begin{vmatrix} x & y & z-1 \\ \sin(\theta_c) & 0 & \cos(\theta_c) - 1 \\ \sin(\theta_c)\cos(\Omega) & \sin(\theta_c)\sin(\Omega) & \cos(\theta_c) - 1 \end{vmatrix} = 0$

Finally after simplification, the first tetrahedron is found to be bounded as follows
$$0 \le y \le \sqrt{G}$$
$$y/\tan(\Omega) \le x(y) \le \sin(\theta_c) - \tan(\Omega/2)y$$
$$\cot(\theta_c)\,[x + \tan(\Omega/2)\,y] \le z(x,y) \le 1 - \tan(\alpha)\cos(\Omega/2)[x + \tan(\Omega/2)\,y]$$

## III. Definition and values of the $\{c_n\}$ constants

In evaluation of Fourier transforms, it is helpful to define some constants

$c_1 = -\tan(\alpha)\cos\left(\frac{\Omega}{2}\right) = -(\sqrt{5} - 1)/2$

$c_2 = \sin(\theta_c)\sin(\Omega) = \sqrt{1/2 + \sqrt{5}/10}$

$c_3 = -\tan(\alpha)\cos\left(\frac{\Omega}{2}\right) - \cot(\theta_c) = -\sqrt{5}/2$

$c_4 = \tan\left(\frac{\Omega}{2}\right) + \frac{1}{\tan(\Omega)} = \sqrt{2 - 2/\sqrt{5}}$

$c_5 = -\cot(\theta_c)/\left[\tan(\alpha)\cos\left(\frac{\Omega}{2}\right)\right] = -(\sqrt{5} + 1)/4$

$c_6 = \cot(\theta_c)\tan\left(\frac{\Omega}{2}\right) = \frac{\sqrt{5 - 2\sqrt{5}}}{2}$

$c_7 = \sin(\theta_c) + \left(\frac{1}{2}\right)(c_1 - \cot(\theta_c))\sin^2(\theta_c) = 1/\sqrt{5}$

$c_8 = -c_4 + \sin(\theta_c)\left(c_1\tan\left(\frac{\Omega}{2}\right) - c_6\right) - \sin(\theta_c)\tan\left(\frac{\Omega}{2}\right)(c_1 - \cot(\theta_c)) = -\sqrt{2 - 2/\sqrt{5}}$

$c_9 = -c_4\left(c_1\tan\left(\frac{\Omega}{2}\right) - c_6\right) + \left(\frac{1}{2}\right)(c_1 - \cot(\theta_c))(\tan^2\left(\frac{\Omega}{2}\right) - 1/\tan^2(\Omega)) = (\sqrt{5} - 1)/2$

$c_{10} = c_1 - \cot(\theta_c) = -\sqrt{5}/2$

$c_{11} = \sin(\theta_c)(c_1 - \cot(\theta_c)) + 1 = -1$

$c_{12} = c_1\cot(\Omega) - \cot(\theta_c)\cot(\Omega) + c_1\tan\left(\frac{\Omega}{2}\right) - c_6 = -0.5\sqrt{10 - 2\sqrt{5}}$

$c_{13} = \tan\left(\frac{\Omega}{2}\right) = \sqrt{5 - 2\sqrt{5}}$

$c_{14} = \cot(\Omega) = \sqrt{1 - 2/\sqrt{5}}$

$c_{15} = \cot(\theta_c) = 0.5$

$c_{16} = c_2 c_4 = 2/\sqrt{5}$

$c_{17} = c_1 c_{13} = -\sqrt{50 - 22\sqrt{5}}/2$

## IV. Indefinite integral lemma

$$\int [ay^2 + by + c]e^{\omega y}\,dy = \frac{e^{\omega y}}{\omega}\left[ay^2 + \left(b - \frac{2a}{\omega}\right)y + \left(c + \frac{2a}{\omega^2} - \frac{b}{\omega}\right)\right] + Const.$$

## V. The function $E(p)$ and its derivatives

It is helpful to define a function $E(x)$ and also explicitly calculate its derivatives, as follows

$$E(p) = \int_{y=0}^{y=c_2} e^{-i2\pi py}\,dy = \begin{cases} c_2 \dfrac{e^{-i2\pi c_2 p} - 1}{-i2\pi c_2 p} & \text{if } p \ne 0 \\ c_2 & \text{if } p = 0 \end{cases}$$

Differentiating $E(p)$ at any nonzero point gives
$$E'(p) = \frac{(-i2\pi c_2)e^{-i2\pi c_2 p}(-i2\pi p) - (-i2\pi)(e^{-i2\pi c_2 p} - 1)}{(-i2\pi p)^2} = \frac{(1 + i2\pi c_2 p)e^{-i2\pi c_2 p} - 1}{i2\pi p^2}$$

Differentiating $E(p)$ at $p = 0$ yields
$$E'(0) = \lim_{p \to 0}\frac{E(p) - E(0)}{p - 0} = -i\pi c_2^2$$

So,

$$E'(p) = \begin{cases} i2\pi c_2^2 \dfrac{(1 + i2\pi c_2 p)e^{-i2\pi c_2 p} - 1}{(i2\pi c_2 p)^2} & if\ p \neq 0 \\ -i\pi c_2^2 & if\ p = 0 \end{cases}$$

And similarly,

$$E''(p) = \begin{cases} -4\pi^2 c_2^3 \dfrac{2 - [1 + (1 + i2\pi c_2 p)^2]\,e^{-i2\pi c_2 p}}{(i2\pi c_2 p)^3} & if\ p \neq 0 \\ -\left(\dfrac{4}{3}\right)\pi^2 c_2^3 & if\ p = 0 \end{cases}$$

On the other hand, differentiating both sides of the equation defining $E(p)$ with respect to the variable $p$ (and assuming interchangability of the integration with respect to $y$ and differentiation with respect to $p$; i.e., uniform convergence), and also using Lemmas 1 and 2, we have

$$E^{(n)}(p) = \frac{d^n E}{dp^n} = \int_{y=0}^{y=c_2} \frac{\partial^{(n)}}{\partial p^n} e^{-i2\pi py} dy = (-i2\pi)^n \int_{y=0}^{y=c_2} y^n e^{-i2\pi py} dy$$

$$\int_{y=0}^{y=c_2} y^n e^{-i2\pi py} dy = E^{(n)}(p)/(-i2\pi)^n$$

$$\int_{y=0}^{y=c_2} y e^{-i2\pi py} dy = \frac{E'(p)}{(-i2\pi)} = \begin{cases} c_2^2 \dfrac{1 - (1 + i2\pi c_2 p)e^{-i2\pi c_2 p}}{(i2\pi c_2 p)^2} & if\ p \neq 0 \\ c_2^2/2 & if\ p = 0 \end{cases}$$

$$\int_{y=0}^{y=c_2} y^2 e^{-i2\pi py} dy = \frac{E''(p)}{(-i2\pi)^2} = \begin{cases} c_2^3 \dfrac{2 - [1 + (1 + i2\pi c_2 p)^2]\,e^{-i2\pi c_2 p}}{(i2\pi c_2 p)^3} & if\ p \neq 0 \\ c_2^3/3 & if\ p = 0 \end{cases}$$

## VI. Auxiliary functions $\eta_n$

The following functions (of the Cartesian components) of the spatial frequency vector $\boldsymbol{q}$ appear in different expressions of the Fourier transform:

$$\eta_0(q_x, q_z) = \frac{-e^{-i2\pi(q_z + \sin(\theta_c)(q_x + c_1 q_z))}}{4\pi^2 q_z (q_x + c_1 q_z)}$$

$$\eta_1(q_x, q_z) = \frac{e^{-i\pi \sin(\theta_c)(2q_x + q_z)}}{2\pi^2 q_z (2q_x + q_z)}$$

$$\eta_2(q_x, q_z) = \frac{e^{-i2\pi q_z}}{4\pi^2 q_z (q_x + c_1 q_z)}$$

$$\eta_3(q_x, q_z) = \frac{-1}{2\pi^2 q_z (2q_x + q_z)}$$

$$\eta_4(q_x, q_z) = c_{14} q_x + + c_1 c_4 q_z$$

$$\eta_5(q_x, q_z) = \cot(\Omega)\, q_x + (c_4/2) q_z$$

$$\eta_6(q_x, q_z) = \frac{e^{-i2\pi \sin(\theta_c) q_x}}{-i2\pi q_x}\left(c_{11} + \frac{c_{10}}{i2\pi q_x}\right)$$

$$\eta_7(q_x) = \frac{1 + \dfrac{c_{10}}{i2\pi q_x}}{i2\pi q_x}$$

## VII. Conventions, disambiguation, and evaluation of the first Fourier integral (over $z$)

*Conventions regarding the Fourier kernel*
1. The convention used for the *sign* of the Fourier exponential is consistent with many engineering textbooks and the programs Matlab and SciPy, but opposite of the convention common in physics textbooks.
2. The spatial frequency $\boldsymbol{q}$ corresponds to a scattering process with input and output "wavevectors" as $\boldsymbol{q} = \boldsymbol{k}_{out} - \boldsymbol{k}_{in}$, where $|\boldsymbol{k}_{in}| = |\boldsymbol{k}_{out}| = 1/\lambda$ (and *not* $2\pi/\lambda$).

*Ambiguities*
Analytical formulation of Fourier transform can give rise to sinc-like ambiguities. To be on the safe/fast side computationally and also to have a closed-form expression (rather than an infinite series) analytically, we treat the calculation of the integrals in such cases separately.

*First integral (over z)*
**Case I:** $q_z = 0$
$$A(q_z = 0) = 1 + c_{10}x + \left(c_1 \tan\left(\frac{\Omega}{2}\right) - c_6\right)y$$

**Case II:** $q_z \neq 0$
$$A(q_z \neq 0) = \frac{1}{-i2\pi q_z}\left[e^{-i2\pi q_z\left(1+c_1 x + c_1 \tan\left(\frac{\Omega}{2}\right)y\right)} - e^{-i2\pi q_z(\cot(\theta_c)x + c_6 y)}\right]$$

## VIII. Evaluation of the second Fourier integral (over $x$)

**Case I:** $q_z = 0$
$$B(q_z = 0) = \int_{x=\frac{y}{\tan(\Omega)}}^{x=\sin(\theta_c)-\tan\left(\frac{\Omega}{2}\right)y} dx\, e^{-i2\pi q_x x}\left[1 + c_{10}x + \left(c_1 \tan\left(\frac{\Omega}{2}\right) - c_6\right)y\right]$$

**Case I.1:** $q_x = q_z = 0$
$$B(q_x = q_z = 0) = [1 + (c_1' - c_6)y](\sin(\theta_c) - c_4 y) + \left(\frac{1}{2}\right)c_{10}\left\{[\sin(\theta_c) - \tan\left(\frac{\Omega}{2}\right)y]^2 - \frac{y^2}{\tan^2(\Omega)}\right\} = c_7 + c_8 y + c_9 y^2$$

**Case I.2:** $q_x \neq 0, q_z = 0$
$$B(q_x \neq 0, q_z = 0) =$$
$$\left\{\frac{e^{-i2\pi x q_x}}{-i2\pi q_x}\left[c_{10}x + 1 + \left(c_1 \tan\left(\frac{\Omega}{2}\right) - c_6\right)y - \frac{c_{10}}{-i2\pi q_x}\right]\right\}_{x=\frac{y}{\tan(\Omega)}}^{x=\sin(\theta_c)-\tan\left(\frac{\Omega}{2}\right)y} =$$
$$\frac{e^{-i2\pi q_x\left(\sin(\theta_c)-\tan\left(\frac{\Omega}{2}\right)y\right)}}{-i2\pi q_x}\left[c_{10}(\sin(\theta_c) - \tan\left(\frac{\Omega}{2}\right)y) + 1 + \left(c_1 \tan\left(\frac{\Omega}{2}\right) - c_6\right)y - \frac{c_{10}}{-i2\pi q_x}\right] -$$
$$\frac{e^{-i2\pi q_x y/\tan(\Omega)}}{-i2\pi q_x}\left[\frac{c_{10}y}{\tan(\Omega)} + 1 + \left(c_1 \tan\left(\frac{\Omega}{2}\right) - c_6\right)y - \frac{c_{10}}{-i2\pi q_x}\right]$$

$B(q_x \neq 0, q_z = 0) =$
$$\frac{e^{-i2\pi \sin(\theta_c)q_x}}{-i2\pi q_x}\left(c_{11} + \frac{c_{10}}{i2\pi q_x}\right)e^{i2\pi \tan\left(\frac{\Omega}{2}\right)q_x y} + \frac{1}{i2\pi q_x}\left[c_{12}y + \left(1 + \frac{c_{10}}{i2\pi q_x}\right)\right]e^{-\frac{i2\pi q_x y}{\tan(\Omega)}}$$

**Case II:** $q_z \neq 0$
$$B(q_z \neq 0) =$$
$$\int_{x=\frac{y}{\tan(\Omega)}}^{x=\sin(\theta_c)-\tan\left(\frac{\Omega}{2}\right)y} dx \frac{e^{-i2\pi q_x x}}{-i2\pi q_z}\left[e^{-i2\pi q_z\left(1+c_1 x + c_1 \tan\left(\frac{\Omega}{2}\right)y\right)} - e^{-i2\pi q_z(\cot(\theta_c)x + c_6 y)}\right]$$

**Case II.1:** $q_z \neq 0, q_x + \cot(\theta_c)q_z \neq 0, q_x + c_1 q_z \neq 0$
$$B(q_z \neq 0, q_x + \cot(\theta_c)q_z \neq 0, q_x + c_1 q_z \neq 0) =$$
$$\frac{e^{-i2\pi q_z\left(1+c_1 \tan\left(\frac{\Omega}{2}\right)y\right)}}{(-i2\pi q_z)(-i2\pi(q_x + c_1 q_z))}\left[e^{-i2\pi(q_x+c_1 q_z)\left(\sin(\theta_c)-\tan\left(\frac{\Omega}{2}\right)y\right)} - e^{-i2\pi \cot(\Omega)(q_x + c_1 q_z)y}\right]$$
$$-\frac{e^{-i2\pi q_z c_6 y}}{(-i2\pi q_z)(-i2\pi(q_x + \cot(\theta_c)q_z))}\left[e^{-i2\pi(q_x+\cot(\theta_c)q_z)\left(\sin(\theta_c)-\tan\left(\frac{\Omega}{2}\right)y\right)} - e^{-i2\pi \cot(\Omega)(q_x + \cot(\theta_c)q_z)y}\right]$$
$B(q_z \neq 0, q_x + \cot(\theta_c)q_z \neq 0, q_x + c_1 q_z \neq 0) =$
$$[\eta_0(q_x, q_z) + \eta_1(q_x, q_z)]e^{i2\pi \tan\left(\frac{\Omega}{2}\right)q_x y} + \eta_2(q_x, q_z)e^{-i2\pi \eta_4(q_x, q_z)y} + \eta_3(q_x, q_z)e^{-i2\pi \eta_5(q_x, q_z)y}$$

**Case II.2:** $q_z \neq 0, q_x + \cot(\theta_c)q_z = 0$
$$B(q_z \neq 0, q_x + \cot(\theta_c)q_z = 0) = \eta_0(q_x, q_z)e^{i2\pi \tan\left(\frac{\Omega}{2}\right)q_x y} + \eta_2(q_x, q_z)e^{-i2\pi \eta_4(q_x, q_z)y} +$$
$$\int_{x=\frac{y}{\tan(\Omega)}}^{x=\sin(\theta_c)-\tan(\Omega/2)y} dx \frac{e^{-i2\pi q_z c_6 y}}{i2\pi q_z}$$

$B(q_z \neq 0, q_x + \cot(\theta_c)q_z = 0) =$
$$\eta_0(q_x, q_z)e^{i2\pi \tan\left(\frac{\Omega}{2}\right)q_x y} + \eta_2(q_x, q_z)e^{-i2\pi \eta_4(q_x, q_z)y} + \frac{e^{-i2\pi q_z c_6 y}}{i2\pi q_z}(\sin(\theta_c) - c_4 y)$$

**Case II.3:** $q_z \neq 0, q_x + c_1 q_z = 0$
$$B(q_z \neq 0, q_x + c_1 q_z = 0) = \eta_1(q_x, q_z)e^{i2\pi \tan\left(\frac{\Omega}{2}\right)q_x y} + \eta_3(q_x, q_z)e^{-i2\pi \eta_5(q_x, q_z)y} +$$
$$\int_{x=\frac{y}{\tan(\Omega)}}^{x=\sin(\theta_c)-\tan\left(\frac{\Omega}{2}\right)y} dx \frac{e^{-i2\pi q_z(1+c_1 \tan(\Omega/2)y)}}{-i2\pi q_z}$$
$B(q_z \neq 0, q_x + c_1 q_z = 0) =$

$$\eta_1(q_x, q_z)e^{i2\pi \tan(\frac{\Omega}{2})q_x y} + \eta_3(q_x, q_z)e^{-i2\pi\eta_5(q_x, q_z)y} + \frac{e^{-i2\pi q_z(1+c_{17}y)}}{-i2\pi q_z}(\sin(\theta_c) - c_4 y)$$

## IX. Evaluation of the third Fourier integral (over $y$)

$$F_{\frac{T1}{20}-S-IH}(\boldsymbol{q}) = \int_{y=0}^{y=c_2} dy\, e^{-i2\pi y q_y} B(y)$$

**Case I:** $q_z = 0$
**Case I.1:** $q_x = q_z = 0$
$$F_{\frac{T1}{20}-S-IH}(\boldsymbol{q}) = \int_{y=0}^{y=c_2} dy\, e^{-i2\pi y q_y}[c_7 + c_8 y + c_9 y^2] = c_7 E(q_y) + \frac{c_8}{(-i2\pi)} E'(q_y) + \frac{c_9}{(-i2\pi)^2} E''(q_y)$$

**Case I.2:** $q_x \neq 0, q_z = 0$
$$F_{\frac{T1}{20}-S-IH}(\boldsymbol{q}) = \int_{y=0}^{y=c_2} dy\, e^{-i2\pi y q_y} \left[ \frac{e^{-i2\pi \sin(\theta_c)q_x}}{-i2\pi q_x}\left(c_{11} + \frac{c_{10}}{i2\pi q_x}\right)e^{i2\pi \tan(\frac{\Omega}{2})q_x y} + \frac{1}{i2\pi q_x}[c_{12}y + \left(1 + \frac{c_{10}}{i2\pi q_x}\right)]e^{-\frac{i2\pi q_x y}{\tan(\Omega)}} \right]$$

$$F_{\frac{T1}{20}-S-IH}(\boldsymbol{q}) = \eta_6(q_x)E(q_y - c_{13}q_x) + \eta_7(q_x)E(q_y + c_{14}q_x) + \frac{c_{12}}{i2\pi q_x(-i2\pi)} E'(q_y + c_{14}q_x)$$

**Case II:** $q_z \neq 0$
**Case II.1:** $q_z \neq 0, q_x + \cot(\theta_c)q_z \neq 0, q_x + c_1 q_z \neq 0$
$$F_{\frac{T1}{20}-S-IH}(\boldsymbol{q}) = \int_{y=0}^{y=c_2} dy\, e^{-i2\pi y q_y} \{[\eta_0(q_x, q_z) + \eta_1(q_x, q_z)]e^{i2\pi \tan(\frac{\Omega}{2})q_x y} + \eta_2(q_x, q_z)e^{-i2\pi\eta_4(q_x, q_z)y}$$
$$+ \eta_3(q_x, q_z)e^{-i2\pi\eta_5(q_x, q_z)y}\}$$

$$F_{\frac{T1}{20}-S-IH}(\boldsymbol{q}) =$$
$$[\eta_0(q_x, q_z) + \eta_1(q_x, q_z)]E(q_y - c_{13}q_x) + \eta_2(q_x, q_z)E(q_y + \eta_4(q_x, q_z)) + \eta_3(q_x, q_z)E(q_y + \eta_5(q_x, q_z))$$

**Case II.2:** $q_z \neq 0, q_x + \cot(\theta_c)q_z = 0$
$$F_{\frac{T1}{20}-S-IH}(\boldsymbol{q}) = \eta_0(q_x, q_z)E(q_y - c_{13}q_x) + \eta_2(q_x, q_z)E(q_y + \eta_4(q_x, q_z)) +$$
$$\int_{y=0}^{y=c_2} dy\, e^{-i2\pi y q_y} \frac{e^{-i2\pi q_z c_6 y}}{i2\pi q_z}(\sin(\theta_c) - c_4 y)$$

$$F_{\frac{T1}{20}-S-IH}(\boldsymbol{q}) =$$
$$\eta_0(q_x, q_z)E(q_y - c_{13}q_x) + \eta_2(q_x, q_z)E(q_y + \eta_4(q_x, q_z)) + \frac{\sin(\theta_c)}{i2\pi q_z} E(q_y + c_6 q_z) - \frac{c_4}{4\pi^2 q_z} E'(q_y + c_6 q_z)$$

**Case II.3:** $q_z \neq 0, q_x + c_1 q_z = 0$
$$F_{\frac{T1}{20}-S-IH}(\boldsymbol{q}) = \eta_1(q_x, q_z)E(q_y - c_{13}q_x) + \eta_3(q_x, q_z)E(q_y + \eta_5(q_x, q_z)) +$$
$$\int_{y=0}^{y=c_2} dy\, e^{-i2\pi y q_y} \frac{e^{-i2\pi q_z(1+c_{17}y)}}{-i2\pi q_z}(\sin(\theta_c) - c_4 y)$$

$$F_{\frac{T1}{20}-S-IH}(\boldsymbol{q}) = \eta_0(q_x, q_z)E(q_y - c_{13}q_x) + \eta_2(q_x, q_z)E(q_y + \eta_4(q_x, q_z))$$
$$+ \frac{e^{-i2\pi q_z}}{-i2\pi q_z}\left[\sin(\theta_c) E(q_y + c_{17}q_z) + \frac{c_4}{i2\pi} E'(q_y + c_{17}q_z)\right]$$

## X. Fourier transform of a solid icosahedron with sphericity (spherical coordinate)

Having solved the Fourier transform of the unit icosahedron (or the first tetrahedron) in the Cartesian coordinate, we can establish an identity by rewriting the Fourier transform in the spherical coordinate:

$$\int_{\phi=0}^{\phi=\Omega} d\phi \int_{\theta=0}^{\theta=\theta_{CA}(\phi)} d\theta \sin(\theta) \int_{r=0}^{r=r_{ico}(\theta,\phi)} dr\, r^2 e^{-i2\pi r q \cos(\gamma)} = F_{\frac{T1}{20}-S-IH}(\boldsymbol{q})$$

In this equation:
- $\gamma$ is the spherical (geodesic) distance between the unit vectors $\vec{r}/r$ and $\vec{q}/q$ with $\cos(\gamma) = \cos(\theta)\cos(\theta_q) + \sin(\theta)\sin(\theta_q)\cos(\phi - \phi_q)$.
- $\theta_{CA}(\phi) = \cos^{-1}((G-1)\cos(\phi - \Omega/2))$ denotes the CA edge of the first tetrahedron.
- $r = r_{ico}(\theta, \phi)$ denotes the equation of the surface of an icosahedron. Over *the first tetrahedron* $r_{ico}(\theta, \phi) = \left[\cos(\theta) + \sin(\theta)\tan(\alpha)\cos\left(\phi - \frac{\pi}{5}\right)\right]^{-1}$.

By combining the contribution of all tetrahedra to the Fourier transform, we will have

$$\int_{\phi=0}^{\phi=2\pi} d\phi \int_{\theta=0}^{\theta=\pi} \sin(\theta)\, d\theta \int_{r=0}^{r=r_{ico}(\theta,\phi)} dr\, r^2 e^{-i2\pi r q \cos(\gamma)} = F_{S-IH}(\boldsymbol{q})$$

The last two integrals simply mean spherical integration over the entire $4\pi$ stradians solid angle *in real-space* ($\Omega_r$). With $d\Omega_r = \sin(\theta)\, d\theta d\phi$, the triple integral can be re-written as

$$\oiint d\Omega_r \int_{r=0}^{r=r_{ico}(\Omega_r)} dr\, r^2 e^{-i2\pi rq\cos(\gamma(\Omega_r))} = F_{S-IH}(\boldsymbol{q})$$

The first integration yields

$$C = \int_{r=0}^{r=r_{ico}(\theta,\phi)} dr\, r^2 e^{-i2\pi rq\cos(\gamma)} = \left(\frac{1}{\omega^3}\right)[e^{\omega r}\{(1-\omega r)^2 + 1\}]_{r=0}^{r=r_{ico}(\theta,\phi)}$$

$$C = \left(\frac{1}{\omega^3}\right)[e^{\omega r_{ico}}\{(1-\omega r_{ico})^2 + 1\} - 2]$$

with $\omega = -i2\pi q\cos(\gamma)$. Multiplication of $e^{\omega r_{ico}}$ by each monomial generates an infinite series. By defining the new variable $h = \omega r_{ico}$, we combine all these series into a single one

$$C = \left(\frac{1}{\omega^3}\right)\left[-2 + (2 - 2h + h^2)\sum_{n=0}^{n=\infty}\frac{h^n}{n!}\right] =$$

$$\left(\frac{1}{\omega^3}\right)\left[-2 + 2\left(1 + h + \sum_{n=0}^{n=\infty}\frac{h^{n+2}}{(n+2)!}\right) - 2\left(h + \sum_{n=0}^{n=\infty}\frac{h^{n+2}}{(n+1)!}\right) + \sum_{n=0}^{n=\infty}\frac{h^{n+2}}{n!}\right] =$$

$$= \left(\frac{h}{\omega}\right)^3 \sum_{n=0}^{n=\infty}\frac{h^n}{(n+3)n!}$$

Rewriting $C$ in terms of the original spherical coordinate variables, we have

$$C(\Omega_r, \Omega_q) = r^3{}_{ico}(\Omega_r)\sum_{n=0}^{n=\infty}\frac{[-i2\pi q\cos(\gamma(\Omega_r, \Omega_q))r_{ico}(\Omega_r)]^n}{(n+3)n!}$$

$$\oiint d\Omega_r\, r^3{}_{ico}(\Omega_r)\sum_{n=0}^{n=\infty}\frac{[-i2\pi q\cos(\gamma(\Omega_r, \Omega_q))r_{ico}(\Omega_r)]^n}{(n+3)n!} = F_{S-IH}(\boldsymbol{q})$$

$$\sum_{n=0}^{n=\infty}\frac{(-i2\pi q)^n}{(n+3)n!}\oiint d\Omega_r \cos^n(\gamma(\Omega_r, \Omega_q))\, r_{ico}^{n+3}(\Omega_r) = F_{S-IH}(\boldsymbol{q})$$

Repeating the above procedure for purely radial deformations of an icosahedron will simply change the function $r_{ico}(\Omega_r)$. For a sphere with unity radius, one can write $F_{Sph}(\boldsymbol{q}) = \frac{\sin(2\pi q) - (2\pi q)\cos(2\pi q)}{(2\pi q)^3}$ (remember that in the scaled coordinate, the radius of the circumscribed sphere of the icosahedron is also $R_{ico} = 1$). Using the scaling property of 3D Fourier transform, the Fourier transform of an arbitrary sphere with radius R (note that $R_{ico} = 1$) is $R^3 F_{Sph}(R\boldsymbol{q})$, and one one can write

$$\sum_{n=0}^{n=\infty}\frac{(-i2\pi q)^n}{(n+3)n!}\oiint d\Omega_r \cos^n(\gamma(\Omega_r, \Omega_q))\, R^{n+3} = R^3 F_{Sph}(R\boldsymbol{q})$$

For $R = 1$,

$$\sum_{n=0}^{n=\infty}\frac{(-i2\pi q)^n}{(n+3)n!}\oiint d\Omega_r \cos^n(\gamma(\Omega_r, \Omega_q)) = F_{Sph}(\boldsymbol{q}) = \frac{\sin(2\pi q) - (2\pi q)\cos(2\pi q)}{(2\pi q)^3}$$

$$= \frac{1}{(2\pi q)^3}\sum_{n=0}^{n=\infty}\frac{(-1)^n(2\pi q)^{2n+1}}{(2n+1)!} - \frac{(2\pi q)}{(2\pi q)^3}\sum_{n=0}^{n=\infty}\frac{(-1)^n(2\pi q)^{2n}}{(2n)!} = 2\sum_{n=0}^{n=\infty}\frac{(-i2\pi q)^{2n}}{(2n+3)!}(n+1)$$

Matching the coefficients of the *identity* results in the following expression for even $n$. For odd $n$, the integral simply vanishes.

$$\oiint d\Omega_r \cos^{2n}(\gamma(\Omega_r, \Omega_q)) = \frac{1}{(2n+1)}$$

For a spherically-deformed icosahedron with sphericitry of $\sigma$

$$F_{S-Def-IH}(q) = \sum_{n=0}^{n=\infty}\frac{(-i2\pi q)^n}{(n+3)n!}\oiint d\Omega_r \cos^n(\gamma(\Omega_r, \Omega_q))[(1-\sigma)r_{ico}(\Omega_r) + \sigma]^{n+3}$$

We have been able to formulate the Fourier transform, and also evaluate it and achieve an analytical expression (as an infinite series, but not a closed-form expression) for it. We note that both terms in the integrand are smaller than (or at most equal to) unity and decay quickly as $n$ increases. The factorial term in the denominator of the expansion also results in a quick decay of higher-order terms. So, in practice, the series can be truncated with just a few terms. The integral has also a clear closed boundary, and there are special quadrature rules (S.39) for numerical calculation of such integrals.

## XI. Expansion and integration of the term $\cos(\gamma(\Omega_r, \Omega_q))^n$

Surface integrals encountered in Section X of *Supplementary Materials* can be evaluated by expanding the term $\cos^n(\gamma)$

$$\cos(\gamma(\Omega_r, \Omega_q))^n = [\cos(\theta)\cos(\theta_q) + \sin(\theta)\sin(\theta_q)\cos(\phi - \phi_q)]^n =$$

$$\sum_{l=0}^{n} C_n^l \sin^l(\theta)\sin^l(\theta_q)\cos^l(\phi - \phi_q)\cos^{n-l}(\theta)\cos^{n-l}(\theta_q)$$

We also note that for any non-negative integer $l$, $\cos^l(\phi - \phi_q)$ can be written as follows

$$\cos^l(\phi - \phi_q) = \sum_{m=0}^{l} e_m \cos\left(m(\phi - \phi_q)\right)$$

where a coefficient $e_m$ is nonzero only for even or only for odd values of $m$ (corresponding to even and odd values of $l$, respectively). Considering the following identity

$$\cos^l(x) = \text{Real}\left\{2^{-l}[e^{ix} + e^{-ix}]^l\right\} = 2^{-l}\text{Real}\left\{\sum_{s=0}^{l} C_l^s e^{isx} e^{-i(l-s)x}\right\} = 2^{-l}\sum_{s=0}^{l} C_l^s \cos[(l-2s)x]$$

a non-zero coefficient $e_m$ originates from the terms for which $|l - 2s| = m$ or $s = (l \mp m)/2$. With the value of $e_m$ taken care of, we proceed with the main derivation as

$$\cos\left(\gamma(\Omega_r, \Omega_q)\right)^n = \sum_{l=0}^{n}\sum_{m=0}^{l} e_m C_n^l \sin^l(\theta) \sin^l(\theta_q) \cos^{n-l}(\theta) \cos^{n-l}(\theta_q) \cos\left(m(\phi - \phi_q)\right)$$

Using the above equation and recalling $r_{ico}(\theta, \phi) = \left[\cos(\theta) + \sin(\theta)\tan(\alpha)\cos\left(\phi - \frac{\pi}{5}\right)\right]^{-1}$ over the first tetrahedron, we can rewrite the integral in the expansion of $F_{S-Def-T1}(\mathbf{q})$, as

$$I_n^{T1} = \iint d\Omega_r \cos^n\left(\gamma(\Omega_r, \Omega_q)\right)\left[(1-\sigma)r_{ico}(\Omega_r) + \sigma\right]^{n+3} =$$

$$\int_{\phi=0}^{\phi=\Omega} d\phi \int_{\theta=0}^{\theta=\theta_{CA}(\phi)} d\theta \sin(\theta) \sum_{l=0}^{n}\sum_{m=0}^{l} e_m C_n^l \sin^l(\theta) \sin^l(\theta_q) \cos^{n-l}(\theta) \cos^{n-l}(\theta_q) \cos\left(m(\phi - \phi_q)\right)\left[\frac{(1-\sigma)}{\left[\cos(\theta) + \sin(\theta)\tan(\alpha)\cos\left(\phi - \frac{\pi}{5}\right)\right]} + \sigma\right]^{n+3} =$$

$$\sum_{l=0}^{n} C_n^l \sum_{m=0}^{l} e_m \int_{\phi=0}^{\phi=\Omega} d\phi \int_{\theta=0}^{\theta=\theta_{CA}(\phi)} d\theta \sin(\theta) \sin^l(\theta) \sin^l(\theta_q) \cos^{n-l}(\theta) \cos^{n-l}(\theta_q) \cos\left(m(\phi - \phi_q)\right)\left[\frac{(1-\sigma)}{\left[\cos(\theta) + \sin(\theta)\tan(\alpha)\cos\left(\phi - \frac{\pi}{5}\right)\right]} + \sigma\right]^{n+3} =$$

$$\sum_{l=0}^{n} C_n^l \sin^l(\theta_q) \cos^{n-l}(\theta_q) \sum_{m=0}^{l} e_m \int_{\phi=0}^{\phi=\Omega} d\phi \int_{\theta=0}^{\theta=\theta_{CA}(\phi)} d\theta \sin(\theta) \sin^l(\theta) \cos^{n-l}(\theta) \left[\frac{(1-\sigma)}{\left[\cos(\theta) + \sin(\theta)\tan(\alpha)\cos\left(\phi - \frac{\pi}{5}\right)\right]} + \sigma\right]^{n+3} [\cos(m\phi)\cos(m\phi_q) + \sin(m\phi)\sin(m\phi_q)] =$$

$$\sum_{l=0}^{n} C_n^l \sin^l(\theta_q) \cos^{n-l}(\theta_q) \sum_{m=0}^{l} e_m [\cos(m\phi_q) I_{n,l,m}^c + \sin(m\phi_q) I_{n,l,m}^s]$$

So, the Fourier transform of the first spherically-deformed tetrahedron can be written as an analytical infinite series of $\mathbf{q}$, with no numerical integration involving $\mathbf{q}$ components:

$$F_{S-Def-T1}(\mathbf{q}) = \sum_{n=0}^{\infty} \frac{(-i2\pi q)^n}{(n+3)n!} \sum_{l=0}^{n} C_n^l \sin^l(\theta_q) \cos^{n-l}(\theta_q) \sum_{m=0}^{l} e_m [\cos(m\phi_q) I_{n,l,m}^c + \sin(\phi_q) I_{n,l,m}^s]$$

The two discrete arrays $I_{n,l,m}^s$ and $I_{n,l,m}^s$ are independent of $\mathbf{q}$. They can be calculated in a one-time-only problem with arbitrarily high accuracy, and used later with arbitrary choices of (rotated) $\mathbf{q}$, irrespective of its discretization scheme or density. We can further simplify the above expansion by defining new constants, as follows:

$$I_{n,l,m}^t = \frac{(2\pi)^n}{(n+3)n!} C_n^l e_m \sqrt{(I_{n,l,m}^c)^2 + (I_{n,l,m}^s)^2}$$

$$\Phi_{n,l,m}^t = \text{atan2}(I_{n,l,m}^c, I_{n,l,m}^s)$$

$$F_{S-Def-T1}(\mathbf{q}) = \sum_{n=0}^{\infty} (-iq)^n \sum_{l=0}^{n} \sin^l(\theta_q) \cos^{n-l}(\theta_q) \sum_{m=0}^{l} I_{n,l,m}^t \sin(m\phi_q + \Phi_{n,l,m}^t)$$

Note that the above expression is indeed a (real) spherical harmonic expansion in the spherical coordinate representation of the reciprocal space.

Since the final Fourier transform of the icosahedron with inversion symmetry is only a real function, one can establish a spherical coordinate identity, and also simplify the calculation of Fourier transform by reducing the complex series to a real one with faster-decaying terms. It also reduces the size of the two arrays $I_{n,l,m}^s$ and $I_{n,l,m}^s$ by a factor of 2, as only even-order terms ($n = 0,2,4,...$) need to be calculated:

$$F_{S-Def-T1}^{Real}(\mathbf{q}) = \sum_{n=0}^{\infty} (-q^2)^n \sum_{l=0}^{2n} \sin^l(\theta_q) \cos^{2n-l}(\theta_q) \sum_{m=0}^{l} I_{2n,l,m}^t \sin(m\phi_q + \Phi_{2n,l,m}^t)$$

$$F_{S-Def-T1}^{imag}(\mathbf{q}) = -q\sum_{n=0}^{\infty} (-q^2)^n \sum_{l=0}^{2n+1} \sin^l(\theta_q) \cos^{2n-l+1}(\theta_q) \sum_{m=0}^{l} I_{2n+1,l,m}^t \sin(m\phi_q + \Phi_{2n+1,l,m}^t)$$

$$F_{S-Def}(\boldsymbol{q}) = 2\sum_{p=1}^{10} F_{S-Def-T1}^{Real}(R_p^{-1}\boldsymbol{q})$$

$$\sum_{p=1}^{10} F_{S-Def-T1}^{imag}(R_p^{-1}\boldsymbol{q}) = 0$$

The above expression can be calculated for the extreme cases of $\sigma = 1$ and $\sigma = 0$ and intermediate values as

$$F_{S-Def-IH}(\boldsymbol{q})|_{\sigma=0} = \sum_{n=0}^{n=\infty} \frac{(-i2\pi q)^n}{(n+3)n!} \oiint d\Omega_r \cos^n(\gamma(\Omega_r,\Omega_q)) r_{ico}^{n+3}(\Omega_r) = F_{S-IH}(\boldsymbol{q})$$

$$F_{S-Def-IH}(\boldsymbol{q})|_{\sigma=1} = \sum_{n=0}^{n=\infty} \frac{(-i2\pi q)^n}{(n+3)n!} \oiint d\Omega_r \cos^n(\gamma(\Omega_r,\Omega_q)) = F_{Sph}(\boldsymbol{q}) = \sum_{n=0}^{n=\infty} \frac{2(n+1)(-i2\pi q)^{2n}}{(2n+3)!}$$

$$F_{S-Def-IH}(\boldsymbol{q})|_{0<\sigma<1} = 2\sum_{p=1}^{10} F_{S-Def-T1}^{Real}(R_p^{-1}\boldsymbol{q})$$

$$F_{S-Def-T1}^{Real}(\boldsymbol{q}) = \sum_{n=0}^{\infty}(-q^2)^n \sum_{l=0}^{2n} \sin^l(\theta_q)\cos^{2n-l}(\theta_q) \sum_{m=0}^{l} I_{2n,l,m}^t \sin(m\phi_q + \Phi_{2n,l,m}^t)$$

Furthermore, the double-integrals encountered in the calculation of $I_{n,l,m}^s$ and $I_{n,l,m}^c$ can be simplified by calculating the first one analytically, as shown below. To avoid similar derivations, we define the index $h \in \{s, c\}$, and $f_m^h(\phi) = \cos(m\phi)$ for $h = c$ and $f_m^h(\phi) = \sin(m\phi)$ for $h = s$.

$$I_{n,l,m}^h = \int_{\phi=0}^{\phi=\Omega} d\phi f_m^h(\phi) \int_{\theta=0}^{\theta=\theta_{CA}(\phi)} d\theta \sin(\theta)\sin^l(\theta)\cos^{n-l}(\theta)\left[\frac{(1-\sigma)}{\cos(\theta)+\sin(\theta)\tan(\alpha)\cos\left(\phi-\frac{\pi}{5}\right)} + \sigma\right]^{n+3}$$

By defining the *parameter* $\epsilon_\phi = \tan(\alpha)\cos\left(\phi - \frac{\pi}{5}\right)$ and change of the polar variable as $z = \tan(\theta/2)$, we have

$$J_{n,l}(\phi) = \int_{\theta=0}^{\theta=\theta_{CA}(\phi)} d\theta \sin(\theta)\sin^l(\theta)\cos^{n-l}(\theta)\left[\frac{(1-\sigma)}{\cos(\theta)+\sin(\theta)\tan(\alpha)\cos\left(\phi-\frac{\pi}{5}\right)} + \sigma\right]^{n+3}$$

$$J_{n,l}(\phi) = \int_0^{\tan(\theta_{CA}(\phi)/2)} \frac{2dz}{1+z^2}\left(\frac{2z}{1+z^2}\right)^{l+1}\left(\frac{1-z^2}{1+z^2}\right)^{n-l}\left[\frac{(1-\sigma)}{\left(\frac{1-z^2}{1+z^2}\right)+\epsilon_\phi\left(\frac{2z}{1+z^2}\right)} + \sigma\right]^{n+3}$$

For a given $\phi$, the integral over $\theta$ can be calculated analytically, as the (partial fraction) decomposition and integration of such an integral of a *real* variable is straightforward (Silverman, 1985).

## XII. FM-like modulation of icosahedral harmonics

The product of two icosahedral harmonics $J_{l1}(\Omega)$ and $J_{l2}(\Omega)$ inherits both icosahedral symmetry and axes of symmetry from them. For simplicity, we assume that not only the original icosahedral harmonics, but also their product can be expressed using low-order non-degenerate icosahedral harmonics. Extension to higher orders follows a similar procedure by including an additional degeneracy parameter. By expanding functions into icosahedral and corresponding spherical spectra, we can write:

$$J_{l1}(\Omega)J_{l2}(\Omega) = \sum_{m1=-l1}^{l1} \sum_{m2=-l2}^{l2} b_{l1,m1}b_{l2,m2}Y_{l1,m1}(\Omega)Y_{l2,m2}(\Omega)$$

$$J_{l1}(\Omega)J_{l2}(\Omega) = \sum_{l3} D_{l3}J_{l3}(\Omega) = \sum_{l3} D_{l3} \sum_{m3=-l3}^{l3} b_{l3,m3}Y_{l3,m3}(\Omega)$$

We simply need to equate the right-hand sides of these two equations and find the projections of different terms on a given spherical harmonic. We multiply both series by $Y_{l,m}(\Omega)$ (for a given $l, m$ with nonzero $b_{l,m}$) and integrate over the entire span of $\Omega$. With the first series, the integrals over the product of three spherical harmonics have known values; i.e., Clebsch–Gordan coefficients $C_{m1,m2,m}^{l1,l2,l}$. With the second series, all spherical harmonic overlap integrals vanish, except for $Y_{l,-m}(\Omega)$, which has a leading coefficient of $D_l b_{l,-m}$. The unknown coefficients $D_l$ are simply found to be

$$D_l = \sum_{m1=-l1}^{l1} \sum_{m2=-l2}^{l2} (b_{l1,m1}b_{l2,m2}/b_{l,-m})C_{m1,m2,m}^{l1,l2,l}$$

The indices $l$ and $m$ run over spherical harmonic indices with nonzero contribution to icosahedral harmonics. This formula represents a set of redundant equations corresponding to different values of $m$ generating the same value for the sought unknown $D_l$. For convenience, we choose $m = 0$ (common for all selection-allowed values of $l$). We also rewrite $D_l$ as $D_{l,l_1,l_2}$ to emphasize the dependence on $l1$ and $l2$. Finally:

$$J_{l1}(\Omega)J_{l2}(\Omega) = \sum_l J_l(\Omega)D_{l,l_1,l_2} = \sum_l J_l(\Omega)\left[\sum_{m1=-l1}^{l1}\sum_{m2=-l2}^{l2}(b_{l1,m1}b_{l2,m2}/b_{l,0})C_{m1,m2,0}^{l1,l2,l}\right]$$

# XIII. Non-icosahedral functions

Orthogonality of a non-icosahedral function and any non-degenerate icosahedral harmonic of order $l_0$ requires zero overlap of their spherical patterns:

$$f(\Omega) = f_{ico}(\Omega) + f_{non-ico}(\Omega) = f_{ico}(\Omega) + \sum_{l=0}^{\infty}\sum_{m=-l}^{l} a_{l,m} Y_{lm}(\Omega)$$

$$\oiint d\Omega J_{l_0}(\Omega) \sum_{l=0}^{\infty}\sum_{m=-l}^{l} a_{l,m} Y_{lm}(\Omega) = \oiint d\Omega \sum_{m'=-l_0}^{l_0} b_{l_0,m'} Y_{l_0 m'}(\Omega) \sum_{l=0}^{\infty}\sum_{m=-l}^{l} a_{l,m} Y_{lm}(\Omega) =$$

$$\sum_{l=0}^{\infty}\sum_{m=-l}^{l}\sum_{m'=-l_0}^{l_0} b_{l_0,m'} a_{l,m} \oiint d\Omega Y_{lm}(\Omega) Y_{l_0 m'}(\Omega) = \sum_{l=0}^{\infty}\sum_{m=-l}^{l}\sum_{m'=-l_0}^{l_0} b_{l_0,m'} a_{l,m} \delta_{l,l_0} \delta_{m,-m'} = \sum_{m'=-l_0}^{l_0} b_{l_0,m'} a_{l_0,-m'}$$

$$= \sum_{m'=-l_0}^{l_0} b^*_{l_0,-m'} a_{l_0,-m'} = \boldsymbol{a}_{l_0} \cdot \boldsymbol{b}^*_{l_0}$$

# XIV. Formulation of reflection off spherical and cylindrical objects

*Reflection from a spherical hydration shell*

A sphere can be modeled as $\vec{r}_s = \vec{C} + R[\sin(\theta)\cos(\varphi), \sin(\theta)\sin(\varphi), \cos(\theta)]$, where $\vec{C}$ and $R$ represent the center and the radius of the sphere, $\vec{r}_s$ is an arbitrary point on the sphere, and $(\theta, \varphi)$ are azimuth and zenith angles. The parameter $t$ can be deleted in this alternative implicit equation: $(\vec{r}_s - \vec{C}) \cdot (\vec{r}_s - \vec{C}) = R^2$. The intersection of the sphere with a line $\vec{r}_l = \vec{P}_0 + t * \hat{d}$ (if any), occurs when a point satisfies both equations:
$(\vec{P}_0 + t * \hat{d} - \vec{C}) \cdot (\vec{P}_0 + t * \hat{d} - \vec{C}) = R^2$, or
$t^2 - 2[\hat{d} \cdot (\vec{P}_0 - \vec{C})]t + [(\vec{P}_0 - \vec{C}) \cdot (\vec{P}_0 - \vec{C}) - R^2] = 0$

Defining $q_1 = \hat{d} \cdot (\vec{P}_0 - \vec{C})$ and $q_2 = [(\vec{P}_0 - \vec{C}) \cdot (\vec{P}_0 - \vec{C}) - R^2]$, the $t$ values associated with the entrance and exit points of the line (into and out of the sphere) are

$$t_{entrance} = q_1 - \sqrt{q_1^2 - q_2}, t_{exit} = q_1 + \sqrt{q_1^2 - q_2}$$

For $q_1^2 - q_2 < 0$, there is no point of intersection, and for $q_1^2 - q_2 = 0$, the ray is tangential to the sphere. In either case, the ray propagates unaffected. For $q_1^2 - q_2 > 0$, there is only one point of intersection *relevant to the incident beam*, corresponding to $t_{entrance}$; i.e., $\vec{P}_1 = \vec{P}_0 + (q_1 - \sqrt{q_1^2 - q_2}) * \hat{d}$.

According to the law of reflection, 1) the beam reflected off the sphere and the beam transmitted into the sphere both lie in the plane formed by the incident beam and normal to the sphere, and 2) the angles of incidence and reflection (with respect to the surface normal) are the same. So, the unit vector in the direction of the reflected beam is $\hat{d}_{refl} = s_1 \hat{d} + s_2 \hat{n}_1$, where $\hat{n}_1 = (\vec{P}_1 - \vec{C})/|\vec{P}_1 - \vec{C}|$ is the *outward*-pointing unit normal vector at the point of intersection, and $s_1$ and $s_1$ are constants to be found. Multiplying both sides of the equation for $\hat{d}_{refl}$ once by $\hat{d}$ and once by $\hat{n}_1$ and enforcing the second law of reflection ($\hat{d}_{refl} \cdot \hat{d} = \cos(\pi - 2\theta_{inc})$ and $\hat{d}_{refl} \cdot \hat{n}_1 = \cos(\theta_{inc})$, where $\theta_{inc} = \cos^{-1}(-\hat{d} \cdot \hat{n}_1)$ is the incidence angle), one can write $\hat{d}_{refl} = \hat{d} + 2\cos(\theta_{inc})\hat{n}_1$ or

$$\hat{d}_{refl} = \hat{d} - 2(\hat{d} \cdot \hat{n}_1)\hat{n}_1$$

Finally, the reflected ray (carrying the full power of the incident ray) is described by

$$\vec{r}_{l_{refl}} = \vec{P}_1 + t_{refl}\hat{d}_{refl}$$

*Reflection from a cylindrical liquid jet*

Consider a cylinder with the unit vector $\hat{d}_c$ along its axis and (in a known direction), and an arbitrary direction $\hat{a}$ different from $\hat{d}$. We can define an orthogonal coordinate with the two basis $\hat{e} = [\hat{d} \times \hat{a}]/|\hat{d} \times \hat{a}|$ and $\hat{f} = \hat{d} \times \hat{e}$ in perpendicular cross sections of the cylinder. Assuming an arbitrary point $H_0$ on the axis, one can parameterize a point on the axis as $\vec{r}_{axis} = \vec{H}_0 + \alpha R \hat{d}_c$, where $\alpha$ is a unitless and dimensionless parameter, and $R$ is the radius of the cylinder (for scaling purposes). The 1-bit ambiguity regarding the direction of $\hat{d}$ is coupled to (removed by) a similar ambiguity regarding the sign of the parameter $\alpha$.

An arbitrary point on the surface of the cylinder can be parameterized as $\vec{r}_{cyl} = \overrightarrow{OM} = \overrightarrow{OH} + \overrightarrow{HM} = \vec{H}_0 + \alpha R \hat{d}_c + R[\cos(\beta)\hat{e} + \sin(\beta)\hat{f}]$. Note that although $\hat{d}_c - \hat{e} - \hat{f}$ is a complete right-handed orthogonal system in 3D, only two independent directions are needed to generate its 3 vectors. At the end:

$$\vec{r}_{cyl} = \vec{H}_0 + R[\alpha \hat{d}_c + \cos(\beta)\hat{e} + \sin(\beta)\hat{f}]$$

The intersection point of a ray (line) $\vec{r}_l = \vec{P}_0 + t * \hat{d}_l$ with this cylinder should satisfy the equation $\vec{r}_{cyl} = \vec{r}_l$, or alternatively

$$\overrightarrow{P_0 H_0} + R[\alpha \hat{d}_c + \cos(\beta)\hat{e} + \sin(\beta)\hat{f}] - t * \hat{d}_l = 0$$

This is a system of 3 nonlinear equations in terms of 3 unknowns ($\alpha,\beta,t$), or alternatively 3 linear equations in terms of 4 unknowns ($\alpha,\cos(\beta), \sin(\beta),t$) and a nonlinear constraint ($\cos^2(\beta) + \sin^2(\beta) = 1$). There are different ways of choosing 3 *variables* (to solve the matrix equation) and a *parameter*. Choosing $\begin{bmatrix} \alpha \\ \cos(\beta) \\ \sin(\beta) \end{bmatrix}$ as the variable has the benefit of three independent directions with no singularity problem in inversion of the matrix.

$$[\hat{d}_c \quad \hat{e} \quad \hat{f}] \begin{bmatrix} \alpha \\ \cos(\beta) \\ \sin(\beta) \end{bmatrix} = \frac{1}{R}\left(-\overrightarrow{P_0 H_0} + \hat{d}_l t\right)$$

$$\begin{bmatrix} \alpha \\ \cos(\beta) \\ \sin(\beta) \end{bmatrix} = \frac{1}{R}[\hat{d}_c \quad \hat{e} \quad \hat{f}]^{-1}\left(-\overrightarrow{P_0 H_0} + \hat{d}_l t\right) = \vec{u} + \vec{v} t$$

Note that $[\hat{d}_c \quad \hat{e} \quad \hat{f}]$ is known *not* to be singular. The obtained variables will be dependent and parameterized in terms of the fourth one; i.e., $t$. By imposing the constraint $\cos(\beta)^2 + \sin(\beta)^2 = 1$ and knowing the (one-side) limited range of $t$, the unique value of $t$ (if any) can be determined as

$$[u_2 + v_2 t]^2 + [u_3 + v_3 t]^2 = 1$$

$$t = -[u_2 v_2 + u_3 v_3] \pm \sqrt{v_2^2 + v_3^2 - (v_2 u_3 - v_3 u_2)^2}$$

For a non-tangential intersection, there are two valid solutions to the *geometrical* problem, but only the "closer" point is related to the *optical* problem.

## XV. Supplementary References

S38. Zheng, Y. 1994. *http://docs.lib.purdue.edu/cgi/viewcontent.cgi?article=1206&context=ecetr*.
S39. Lebedev, V. I. 1977. Spherical quadrature formulas exact to orders 25–29. Siberian Math. J. 18:99-107.